\documentclass{aa}  

\def\apjl{ApJ}
\def\apjs{ApJS}
\def\ao{Appl.~Opt.}

\def\aap{A\&A}

\usepackage{newtxtext,newtxmath}

\usepackage[T1]{fontenc}
\usepackage{ae,aecompl}
\usepackage{xfrac}
\usepackage{nicefrac}

\usepackage{graphicx}	
\usepackage{amsmath}	
\usepackage{amssymb}	
\usepackage{multirow}
\makeatletter
\newcommand\footnoteref[1]{\protected@xdef\@thefnmark{\ref{#1}}\@footnotemark}
\makeatother
\usepackage{threeparttable}
\usepackage{color}
\usepackage{txfonts}
\usepackage{url}

\def\degree{\mbox{$^{\circ}$}}
\usepackage{mathtools}
\usepackage{appendix}

\DeclarePairedDelimiter\abs{\lvert}{\rvert}
\bibpunct{(}{)}{;}{a}{}{,} 
\begin{document}
\titlerunning{Thick disk origin of FCC\,170}
\title{The Fornax\,3D project: unveiling the thick disk origin in FCC\,170: 
signs of accretion?}


\author{F. Pinna\inst{1}\fnmsep\inst{2}\fnmsep\thanks{E-mail: fpinna@iac.es},
J. Falc\'on-Barroso\inst{1}\fnmsep\inst{2},
M. Martig\inst{3},
M. Sarzi\inst{4},
L. Coccato\inst{5},
E. Iodice\inst{6},
E. M. Corsini\inst{7}\fnmsep\inst{8},
P.T. de Zeeuw\inst{9}\fnmsep\inst{10},
D.A. Gadotti\inst{5},
R. Leaman\inst{11},
M. Lyubenova\inst{5},
R.M. McDermid\inst{12},
I. Minchev\inst{13},
L. Morelli\inst{7}\fnmsep\inst{8}\fnmsep\inst{14},
G. van de Ven\inst{5}\fnmsep\inst{11},
S. Viaene\inst{4}\fnmsep\inst{15}
}

\authorrunning{F. Pinna et al.}

\institute{
{Instituto de Astrof\'isica de Canarias, Calle Via L\'actea s/n, E-38200 La Laguna, Tenerife, Spain}
\and
{Depto. Astrof\'isica, Universidad de La Laguna, Calle Astrof\'isico Francisco S\'anchez s/n, E-38206 La Laguna, Tenerife, Spain}
\and
{Astrophysics Research Institute, Liverpool John Moores University, 146 Brownlow Hill, Liverpool L3 5RF, UK}
\and
{Centre for Astrophysics Research, University of Hertfordshire, College Lane, Hatfield AL10 9AB, UK}
\and
{European Southern Observatory, Karl-Schwarzschild-Strasse 2, 85748 Garching bei Muenchen, Germany}
\and
{INAF--Osservatorio Astronomico di Capodimonte, via Moiariello 16, I-80131 Napoli, Italy}
\and 
{Dipartimento di Fisica e Astronomia `G. Galilei', Universit\`a di
Padova, vicolo dell'Osservatorio 3, I-35122 Padova, Italy}
\and 
{INAF--Osservatorio Astronomico di Padova, vicolo
dell'Osservatorio 5, I-35122 Padova, Italy}
\and
{Sterrewacht Leiden, Leiden University, Postbus 9513, 2300 RA Leiden, The Netherlands}
\and
{Max-Planck-Institut fuer extraterrestrische Physik, Giessenbachstrasse, 85741 Garching bei Muenchen, Germany}
\and
{Max-Planck Institut fuer Astronomie, Konigstuhl 17, D-69117 Heidelberg, Germany}
\and
{Department of Physics and Astronomy, Macquarie University, Sydney, NSW 2109, Australia}
\and
{Leibniz-Institut für Astrophysik Potsdam, An der Sternwarte 16, D-14482, Potsdam, Germany}
\and
{Instituto de Astronom\'ia y Ciencias Planetarias, Universidad de Atacama, Copiap\'o, Chile}
\and
{Sterrenkundig Observatorium, Universiteit Gent, Krijgslaan 281, B-9000 Gent, Belgium}
}

\date{Received XXX; accepted YYY}

 
  \abstract
   {We present and discuss the stellar kinematics and populations of the S0 galaxy FCC\,170 (NGC\,1381) 
in the Fornax cluster, using deep MUSE data from the Fornax\,3D survey. 
We show the maps of the first four moments of the stellar line-of-sight velocity distribution and of the mass-weighted mean stellar age, metallicity and [Mg/Fe] abundance ratio. 
The high-quality MUSE stellar kinematic measurements unveil the structure of this massive galaxy: a nuclear disk, a bar seen as a boxy bulge with a clear higher-velocity-dispersion X shape, a fast-rotating and flaring thin disk and a slower rotating thick disk.
Whereas their overall old age makes it difficult to discuss differences in the formation epoch between these components, we find a clear-cut distinction between metal-rich and less [Mg/Fe]-enhanced populations in the thin-disk, boxy-bulge and nuclear disk, and more metal-poor and [Mg/Fe]-enhanced stars in the thick disk.
Located in the densest region of the Fornax cluster, where signs of tidal stripping have been recently found, the evolution of FCC\,170 might have been seriously affected by its environment.
We discuss the possibility of its "pre-processing" in a subgroup before falling into the present-day cluster, which would have shaped this galaxy a long time ago.
The thick disk displays a composite star formation history, as a significant fraction of younger stars co-exist with the main older thick-disk population. The former sub-population is characterized by even lower-metallicity and higher-[Mg/Fe] values, suggesting that these stars formed later and faster in a less chemically evolved satellite, which was subsequently accreted.
Finally, we discuss evidence that metal-rich and less [Mg/Fe]-enhanced stars were brought in the outer parts of the thick disk by the flaring of the thin disk.
}


\maketitle


\section{Introduction}  \label{sec:intro}
Thick disks were discovered, in external galaxies, several decades ago \citep{Burstein1979}. Afterwards, their observation in edge-on galaxies of all morphological types proved 
their ubiquity \citep[e.g.][]{Dalcanton2002, Yoachim2006, Comeron2018}. 
In the Milky Way disk, \citet{Gilmore1983} identified three different populations:
the young thin disk, the old thin disk and the thick disk. 
The thick disk's stars were older, more metal-poor and kinematically hotter with a scale height of $\sim$\,1.5\,kpc. 
They were also more $\alpha$-enhanced than the stars in the thin disk \citep[e.g.][]{Prochaska2000, Cheng2012}, suggesting a different evolution history and timescale.
In external galaxies, when the thick disk is defined by morphology, it appears as a low-surface brightness red envelope with similar color and 
geometry to the Milky Way's thick disk \citep{Dalcanton2002}. 

Little is known about the origin of thick disks.
Two main formation scenarios are advocated to explain their general properties: thick disks were born already dynamically hot at high redshift or they formed from the thickening of a preexisting thin disk, dynamically heated.
In the first case, stars could have been born from dynamically hot gas, as suggested by the increase in gas velocity dispersion with redshift observed by \citet{Wisnioski2015}, and in agreement with the simulation by e.g. \citet{Forbes2012}. \citet{Elmegreen2006} observed disks already thick in spiral galaxies in the Hubble Space Telescope Ultra Deep Field, characterized by giant star-forming clumps, probably produced by gravitational instabilities in a highly turbulent phase. The thin disk would have formed later. 
In \textit{N}-body chemodynamic simulations by \citet{Brook2004}, the thick-disk stars were born
\textit{in situ} at high redshift, 
before the formation of the thin disk, from gas accreted during the chaotic phase of multiple mergers.  
However,
in the Hubble Space Telescope Frontier Fields, an anticorrelation between the mid-plane brightness and the scale height suggested the early presence of two components, 
(i.e. a faint thick disk and a bright thin disk) in clumpy
spiral galaxies at high redshift (up to $z$\,=\,3, \citealt{Elmegreen2017}). 
Another possibility to form an already hot disk is by direct accretion, as proposed by \citet{Abadi2003}.  
In their simulations, thick-disk stars came from the tidal debris of accreted satellites and have, therefore, an extragalactic origin.

A fast dynamical heating of recently born stars would also form quickly a thick disk.
\citet{Bournaud2009} showed that internal clumps in gas-rich young galaxies at high redshift are capable of rapidly scattering disk stars. 
Minor mergers can heat the disk with multiple events over time or in a violent single event \citep
{Toth1992,Quinn1993,Sellwood1998,Benson2004,Kazantzidis2008,Bournaud2009,Villalobos2009,Qu2011, diMatteo2011, Martig2014a,Martig2014b, Pinna2018}.
As pointed out by \citet{Bournaud2009}, the difference between thick disks formed by internal processes 
and those made by minor mergers heating is that the former show a constant scale height, while
the latter flare.
\citet{Minchev2015} showed, however, that while all mono-age stellar populations unavoidably flare, the flares dominate at different radii (the younger,  the further out) in an inside-out disk formation scenario. These nested flares result in a thick disk which does not flare and has large scale length, consistent with observations of external edge-on galaxies.
On the other hand, secular heating might be responsible for a slow thickening of the disk, through scattering 
off giant molecular clouds or black holes present in the dark halos 
\citep[e.g.][]{Spitzer1953,Villumsen1985, Lacey1985, Hanninen2002, Benson2004}. 
\citet{Lacey1985} found that this thickening would be too inefficient to explain the observed thick disks, in agreement with \citet{Minchev2012c}.

For the Milky Way thick disk, numerous works in the literature proposed different scenarios.
Following a morphological definition of the Milky-Way thick disk,
\citet{Dierickx2010} compared the eccentricity distribution of its stars to simulations. 
These observations were compatible with an \textit{in-situ} formation for most stars, related to gas accretion during 
satellite mergers, while accreted stars could explain higher eccentricities.  
\citet{diMatteo2011} recovered these orbital eccentricities by heating the thin disk with minor mergers in direct orbits in their simulations. 
In most of simulations by \citet{Qu2011}, large scale heights are dominated by stars originally coming from the 
primary 
(thin) disk, vertically heated and radially redistributed by minor mergers. 
In some galaxies stars coming from merging satellites dominate.
In this set of simulations, secular processes were unable to reproduce realistic thick disks.
The chemo-orbital properties found by \citet{Liu2012} pointed to
a double origin for the Milky Way thick disk stars. Some of them would have formed through gas-rich mergers at high redshift, whilst others would have migrated from the thin disk.

Most studies on the thick disks in external galaxies are photometry based. 
Imaging has been able to prove the existence and ubiquity of these disks \citep[e.g.][]{Dalcanton2002}, and it has contributed to characterize them. 
\citet{Rejkuba2009} analyzed the color-magnitude diagram for thick disk 
stars of the Milky Way analogue NGC\,891.
They
found a more chemically evolved thick disk with a 
larger dispersion in metallicities than the Milky Way, suggesting a more active accretion history.
\citet{Yoachim2006}, fitting the surface brightness profiles of 34 very late-type edge-on spiral galaxies with two 
disk components, extracted some general properties for them. 
They found an anticorrelation
of the thick-to-thin disk luminosity and mass ratios with circular velocity $V_c$ (i.e. galaxy total mass).
These trends, confirmed by \citet{Comeron2012}
and \citet{Comeron2018},
pointed towards a different thick disk origin for the two mass ranges defined by $V_c$ above and below $120$\,km\,s$^{-1}$.
\citet{Elmegreen2017} explained that this can be due to the dependence of star formation rate on galaxy mass (see also \citealt{Comeron2012,Comeron2014}).
Massive galaxies probably formed quickly at high redshift and are expected to be overall old 
(including the thick and the thin disk, see some examples in \citealt[]{Comeron2016,Kasparova2016}) and more
$\alpha$-enhanced in their thick disks. Low-mass galaxies might be observed just as an $\alpha$-poor thick component.

In this context, spectroscopy is needed to extract the  
stellar kinematics and populations of the thick disks, giving more insights about their properties and 
a better opportunity to understand their origin. 
However, this kind of analysis 
is extremely challenging due to the very low surface brightness of thick disks. 
For this reason, few  spectroscopic studies extracted the stellar kinematics and even fewer the stellar populations.
\citet{Yoachim2005}, \citet{Yoachim2008b} and \citet{Yoachim2008a} studied the stellar kinematics and populations of the thin and the thick disks in late-type galaxies. They used long-slit 
observations in two different positions with respect to the galaxy plane.
\citet{Kasparova2016}, using a full-spectral-fitting technique on long-slit data, studied the stellar populations of three early-type edge-on disk galaxies.
They found a variety of abundances and age distributions, suggesting different formation pictures in different environments.
While long-slit studies have provided valuable information and hints about the thick disk origin, only 
integral field spectroscopy 
allows us to extract stellar kinematics and populations maps, 
and have a global 
view of the galaxies and their disks.
The first study of this kind was done with VIMOS by \citet{Comeron2015}, 
with low spatial resolution 
and a single bin for the thick disk
dominated region.
This was followed by a second bi-dimensional study by \citet{Comeron2016}, 
this time using deeper IFU data from MUSE allowing several bins in the thick disk region.

Even deeper data are required to go further from the midplane into the very faint thick disk with 
larger spatial resolution.
Moving in this direction, 
in this paper we use deep MUSE integral-field observations from the Fornax\,3D
survey \citep{Sarzi2018} to measure the stellar kinematics and stellar-population properties
of the S0 FCC\,170 (NGC\,1381).
We analyze the implications on the origin of the thick disk of this galaxy.
Section~\ref{sec:FCC170} and \ref{sec:obs} give the relevant information about the galaxy and the data set. 
In \S~\ref{methods} we describe the methods used to extract the stellar kinematics and populations. 
In \S~\ref{results} we show the results on the first four moments of the velocity distribution, age, 
metallicity and [Mg/Fe]. 
These results are discussed in \S~\ref{sec:discussion}, and summed up in \S~\ref{conclusions}.

\section{FCC\,170}
\label{sec:FCC170}
FCC\,170 (NGC\,1381) is an 
S0 
galaxy in the Fornax cluster located at a distance of 21.9\,Mpc \citep{Iodice2018} (so that 10\,arcsec in the sky correspond to $\sim$\,1.06\,kpc). 
This bright galaxy 
has an effective radius in the \textit{B}-band of 13.8\,arcsec ($\simeq$\,1.46\,kpc) \citep{Sarzi2018}.
Its maximum circular velocity is about 280\,km\,s$^{-1}$ \citep{Bedregal2006} and its stellar mass has been estimated as $2.25\times 10^{10}$\,M$_{\sun}$ \citep{Iodice2018}.
In the context of thick disk origin, FCC\,170 is therefore expected to behave as a massive galaxy \citep[e.g.][]{Yoachim2006,Yoachim2008a,Comeron2011b,Comeron2012,Comeron2014,Comeron2018}.

It is edge-on \citep[e.g.][]{Bureau2006} and symmetric about the major and minor axes \citep{deCarvalho1987}. It is an object of interest in the field of galaxy formation and evolution due to its complex structure. The prominent box-shaped bulge \citep[e.g.][]{Lutticke2000}, with a central X shape \citep{Bureau2006}, 
was identified as a bar viewed edge-on by \citet{Chung2004} and \citet{Bedregal2006}.
\citet{Williams2011} suggested the presence of a classical bulge and small disky structure in the center. \citet{Comeron2018} confirmed the existence of a thick disk and used two disky components and a central mass concentration to fit the surface-brightness profile. They determined that the thick disk light starts to dominate over the thin disk
at a distance of 10.8\,arcsec ($\simeq$\,1.1\,kpc) from the midplane. 

FCC\,170 has no important line emission in the optical range \citep{Bureau1999}.
This simplifies the extraction of the stellar populations and makes this galaxy very suitable for studying their vertical distribution.
At the same time, the fact that this galaxy does not show any signs of dust or gas is noteworthy and could be related to the peculiar environment where it lives.
FCC\,170 belongs to the Fornax cluster, which is
the second largest concentration of galaxies relatively close to the Milky Way, after Virgo \citep[e.g.][]{Jordan2007}. Much less massive, Fornax has a core $\sim$\,40\% smaller that Virgo, but twice as dense.
FCC\,170 is located in
this core (at a projected distance of 0.42\degree\,from the central galaxy FCC\,213, from \citealt{Iodice2018}), where intra-cluster light (ICL) has been recently found \citep{Iodice2017a}. 
Hence, this work can provide insights about galaxy evolution and thick disks in dense environments.

\section{Observations and data reduction} \label{sec:obs}
The spectroscopic data used in this work were collected as part of the Fornax 3D survey \citep{Sarzi2018}, a magnitude limited survey of galaxies within the virial radius of the Fornax cluster. The main purpose of this project is to improve
our understanding of the formation and evolution of early and late-type galaxies by means of high-quality and deep integral-field spectroscopic measurements. On average, these reach out to 3.5 half-light radii and down to $B$-band surface brightness of 25\,mag\,arcsec$^{-2}$. The deep character of the Fornax\,3D data makes them ideal to study also the faint, thick-disk regions of edge-on galaxies, such as FCC\,170.

Fornax 3D observations are detailed in \citet{Sarzi2018}. They were performed with the Multi Unit Spectroscopic Explorer (MUSE) \citep{Bacon2010}, mounted on the UT\,4 at the Very Large Telescope (VLT). They made use of high-quality 3D spectroscopy with a spatial sampling of $0\farcs 2 \times 0\farcs 2$ in a large field of view of $1\arcmin \times 1\arcmin$ with the Wide-Field Extended mode.
The wavelength range covered from 4650 to 9300\,\AA\ with nominal spectral resolution of 
2.5\,\AA\,(FWHM) at 7000\,\AA\ and spectral sampling of 1.25\,\AA\,pixel$^{-1}$.
The measured spectral resolution was on average 2.8\,\AA\,(FWHM), due to the adverse impact of combining different slightly offset exposures taken at different position angles.
FCC\,170 was observed with two pointings: one in the central region, with single exposures of 720\,s
and a total on-source time of
one hour, and another in the outer part, with single exposures of 600\,s and a total on-source time of two hours.
The position of the two pointings is indicated in Fig.~A.1 of \citet{Sarzi2018}.
Exposures were few arcseconds dithered and rotated by 90$^{\circ}$ in order to
minimize the signature of the 24 MUSE slices on the field of view.
Immediately before or after each science exposure, dedicated sky exposures were carried out in order to perform sky modeling and subtraction on the single spaxels.

The data reduction was performed using the version 1.6.2 of the MUSE reduction pipeline 
\citep{Weilbacher2012, Weilbacher2016}. It included bias and overscan subtraction, flat-fielding correction, wavelength calibration, determination of the line-spread-function, and illumination
correction.
The observation of a spectro-photometric standard star at twilight was necessary to perform the
flux calibration and first-order
correction of the atmospheric telluric features.
The quality of our data after the flux calibration was assessed in \citet{Sarzi2018} for FCC\,167, as a representative case for the entire Fornax\,3D sample. In Fig.\,5 of their paper, they show the residuals of their GandALF fit to the data \citep{Sarzi2006,FalconBarroso2006} using only dust reddening (no multiplicative polynomials), which was a physical expected factor.
The residuals of that fit did not show any systematic structure on long-wavelength scales, but only fluctuations on a wavelength scale of only up to a few hundred \AA, due to either template mismatch or a non-perfect flux calibration. Therefore, even though similar fluctuations might exist in the data for FCC\,170, these would not impact the overall colour of our spectra, so that our present use of multiplicative polynomials would mimic variations introduced by stellar age and metallicity. 
The different exposures were aligned using reference
sources and then combined in a single datacube (see \citealt{Sarzi2018} for more details). \looseness-2

\section{Analysis methods}
\label{methods}
\subsection{Voronoi binning}
From the reduced and combined MUSE cube, we produced a spatially binned MUSE cube using the Python version of the Voronoi 
tessellation software described in \citet{Cappellari2003}\footnote{\label{note2}
\url{
http://www-astro.physics.ox.ac.uk/~mxc/software/}}.
This method provides an adaptive spatial binning to a target signal-to-noise ratio (SNR) per bin, preserving the maximum spatial resolution.
We required a minimum mean SNR of 40 per spatial bin, in the range from 4750 to 5500\,\AA.
We decided to adopt this value with the purpose of having a very good spatial resolution, after testing that compatible results were obtained with larger target SNR (e.g. 80 or 100). 
We set at 1 the minimum accepted SNR per spaxel to be considered, in order to extend the maps to a larger region of the faint thick disk.
The number of the resulting bins was 10344, 
57\% of them with a SNR\,$>$\,40, only 5\% with SNR\,$<$\,35 and none with SNR\,$\le$\,31. 
The Voronoi binned MUSE cube included spectra cut to the wavelength range from 4750 to 5500\,\AA, 
used for the extraction of stellar kinematics and populations.
We decided not to use the full MUSE spectral range to avoid regions where the sky subtraction could have left residuals.
Also, while the used stellar population models (see the next section) cover up to about 7400\,\AA, age sensitivity weakens towards the red and non-alpha sensitive features are included.

\subsection{Full spectrum fitting} \label{sec:ppxf}
We extracted the stellar kinematics and populations of FCC\,170 from full-spectrum fitting, using the Penalized Pixel-Fitting method (pPXF) described in \citet{Cappellari2004} and upgraded in \citet{Cappellari2017}\footnoteref{note2}. Based on a maximum penalized likelihood approach, this method makes use of expansions of the line-of-sight velocity distribution (LOSVD) as Gauss-Hermite series, to fit the observed spectra with templates from a stellar library. 
We used, to fit the MUSE spectra, the [Mg/Fe]-variable version of MILES single stellar population (SSP) models\footnote{\label{note3}Models and their description are available at \url{http://miles.iac.es/}}
based on BaSTI isochrones \citep{Vazdekis2015}.
We assumed a Kroupa Universal IMF, with slope 1.30.
These models cover a wavelength range between 3540 and 7410\,\AA, and are sampled at a spectral resolution of 2.51\,\AA\,(FWHM, \citealt{FalconBarroso2011}).
They include 12 values of total metallicity [M/H]\,=\,[-2.27, +0.40]\,dex, 
with a resolution between 0.14 and 0.48 dex, 53 values of
age between 0.03 and 14.0\,Gyr, 
with a resolution between 0.01 and 0.5\,Gyr,
and two values of [$\alpha$/Fe]\,$\approx$\,[Mg/Fe]\,=\,0.0 (solar abundance) and 0.4\,dex (supersolar).
The SSP models are 1272 in total.
No models of this kind with more than two values of [Mg/Fe] were available.
While this is not optimal, the way how pPXF assigns weights to the models still offers the chance of interpolating between the two [Mg/Fe] values and find intermediate abundances (see \S~\ref{sec:pop_maps} for more details).
For simplicity, in the following sections, with [$\alpha$/Fe] we refer to the [Mg/Fe] abundance, the only abundance we can measure among the $\alpha$ elements.
Each SSP model has a unit total mass (in solar masses), so that mass-weighted results were obtained from pPXF.

\begin{figure}
\scalebox{0.53}{
\includegraphics[scale=0.55]{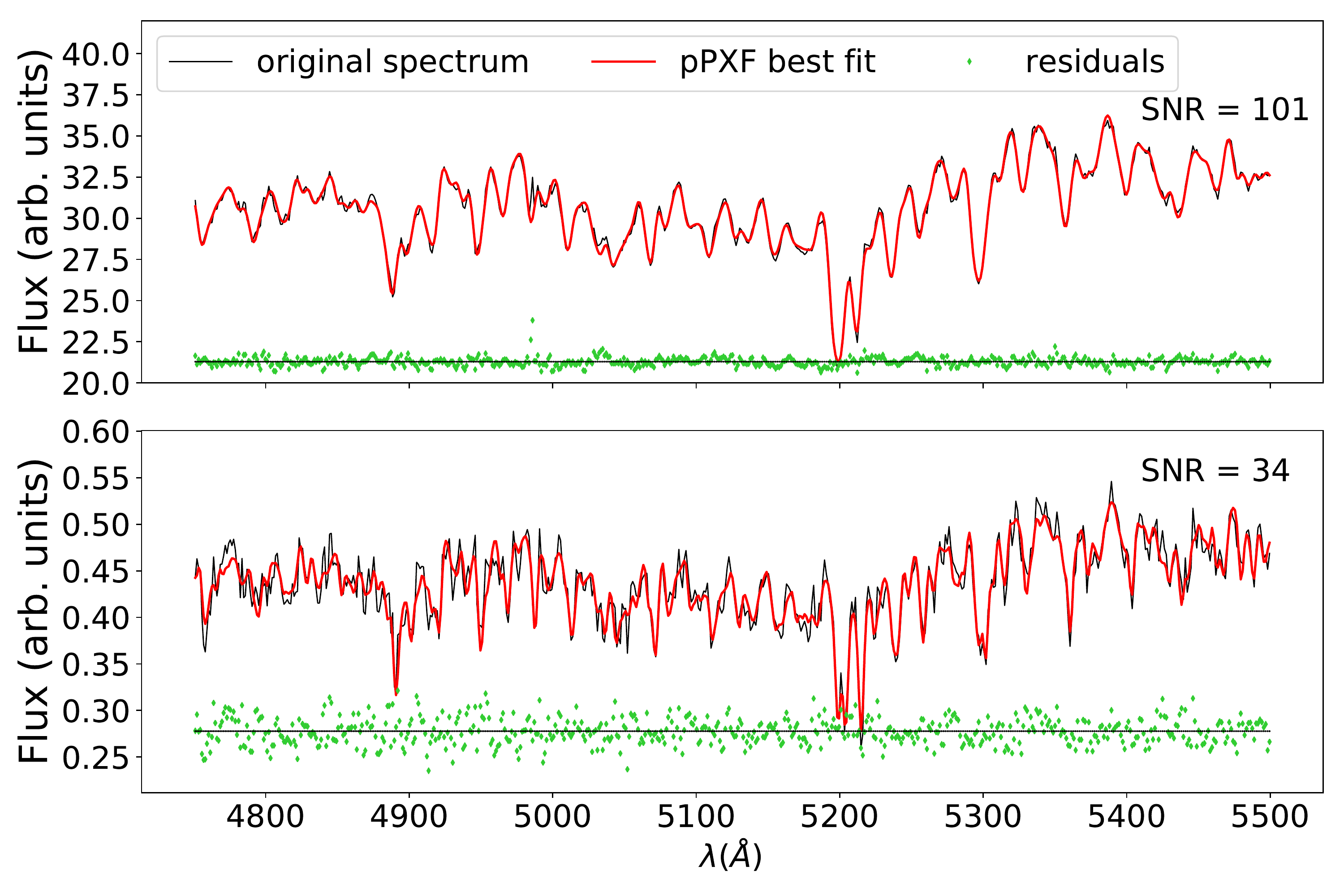}}
\caption{
Examples of two pPXF fits in the wavelength range between 4750 and 5500\,\AA. On the top, the bin with the highest SNR (SNR\,=\,101). On the bottom, a bin with SNR\,=\,34.
The observed spectrum, the best fit and the residuals of the fit are plotted in black, red and green, respectively.
The residuals are arbitrarily shifted for displaying purposes.
} 
\label{fig:fits}
\end{figure}

Since this galaxy has hardly any line emission, it was not necessary to mask any lines when fitting the spectra with pPXF.
To make sure, we compared the best-fit with the original observed spectra and confirmed that no emission lines appeared in the residuals (see examples in Fig.~\ref{fig:fits}). We fitted simultaneously the stellar kinematics and populations, introducing only multiplicative polynomials of 8th order, 
to correct for uncertainties in the spectral calibration. We avoided additive polynomials, which could cause changes in the absorption line strengths and then bias the results in age and metallicity (see also \citealt{Guerou2016}). We verified that the kinematic results were not affected by this approach. 
In particular, to ensure that the velocity dispersion-age degeneracy was not having any influence on the extracted kinematics, we compared the values of velocity dispersions with the ones obtained fitting only the kinematics (with no regularization, see the following paragraph). The differences were in general below the 1\% and showed no spatial correlations in the observed region of the galaxy and no structures along the velocity dispersion range either.

Following \citet{Cappellari2017}, we regularized the stellar-population weights assigned to the SSP models to obtain the best fit to the data. The regularization parameter was chosen by looking for a compromise between smoothing 
(and purging the solutions from potential noise) and not loosing information.
We followed the suggestions given in the same code by the authors as well as in \citet{McDermid2015}
, to calibrate the regularization for the bin with the highest SNR.
First, we rescaled the noise in order to obtain a unit reduced $\chi^2$ with no regularization. Then, we calculated the maximum regularization parameter as the one which increases the $\chi^2$ from the value with no regularization (equal to the number of fitted pixels $N_{good\,pix}$) by the quantity $\Delta\chi^2 \simeq \sqrt{2 N_{good\,pix}}$. This value would be the maximum allowed regularization, leading to the smoothest star formation history (SFH), to give a solution still consistent with the data. 
After that, we performed tests with different regularization parameters between zero and the maximum value, realizing that some peculiar features in the SFH (see \S~\ref{sec:SFH} for details) disappeared above certain smoothing value.
We verified that the position of these maxima in the SFH (i.e. their age, metallicity and [Mg/Fe]) did not change when varying the regularization parameter (see Fig.~\ref{fig:test_comp}).
This smoothed them changing only the mass fraction assigned to them.
Therefore, we decided to select the highest regularization parameter allowing us to see these features, still giving a reasonably smoothed solution.
All the results shown in this work were obtained using a regularization parameter of 0.05, kept constant for all the spatial bins.

In Fig.~\ref{fig:fits} we show two examples of pPXF best fits of spatial bins with high and low SNR.
pPXF gave as output the first four LOSVD moments and the weights given to the individual SSP models to obtain the best fit. 
54 bins were discarded because, despite their SNR being within the limits required during the Voronoi binning, their spectra were not good enough to allow good fits or they were contaminated by foreground stars. 
We kept 10\,290 bins for the analysis in the following sections.

\section{Results}
\label{results}
\subsection{Stellar kinematics}
\begin{figure*}[!h]
\centering
\resizebox{.88\textwidth}{!}
{\includegraphics[scale=1]{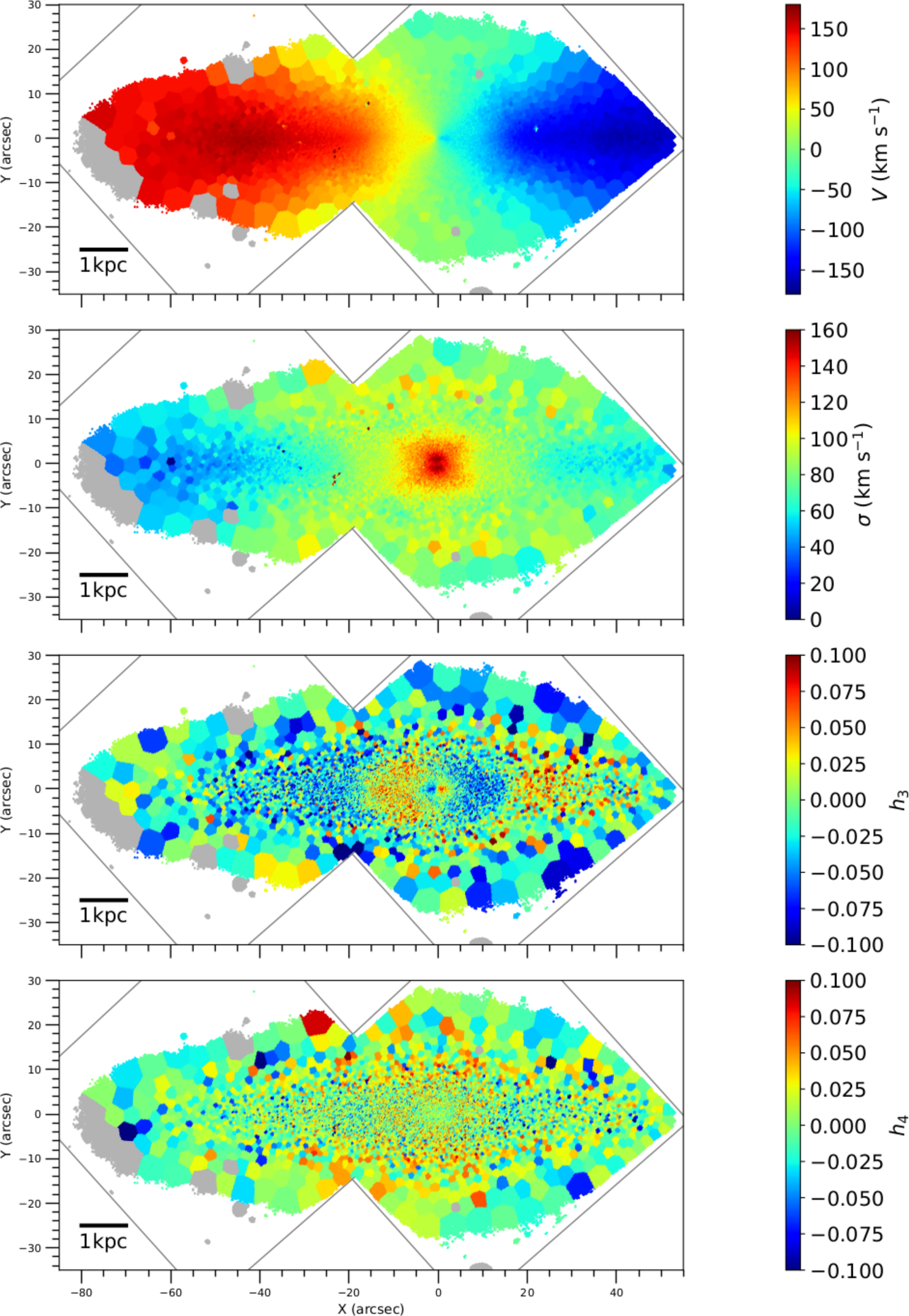}}
\caption{Maps of the first four moments of the stellar LOSVD. From top to bottom: mean velocity $V$, velocity dispersion $\sigma$, third Gauss-Hermite moment $h_3$ (skewness) and fourth Gauss-Hermite moment $h_4$ (kurtosis). The discarded bins and the position of the two MUSE pointings are plotted in grey. A scale bar on bottom-left
of each map indicates the correspondence with physical units.
}
\label{fig:kin}
\end{figure*}

In Fig.~\ref{fig:kin}, we show the maps of mean velocity $V$, velocity dispersion $\sigma$ and third and fourth-order Gauss-Hermite moments, $h_3$ (skewness) and $h_4$ (kurtosis), of the LOSVD.
The discarded bins and the two MUSE pointings are plotted in grey.
In Appendix~\ref{sec:unc}, we give the kinematic uncertainties and we explain how they were computed. The structural components that will be defined in \S~\ref{sec:morph} (Fig.~\ref{fig:morph}) can be distinguished in the kinematic maps described as follows.
\begin{figure*}[!h]
\centering
\resizebox{1.\textwidth}{!}
{
\includegraphics[scale=1,width=\textwidth]{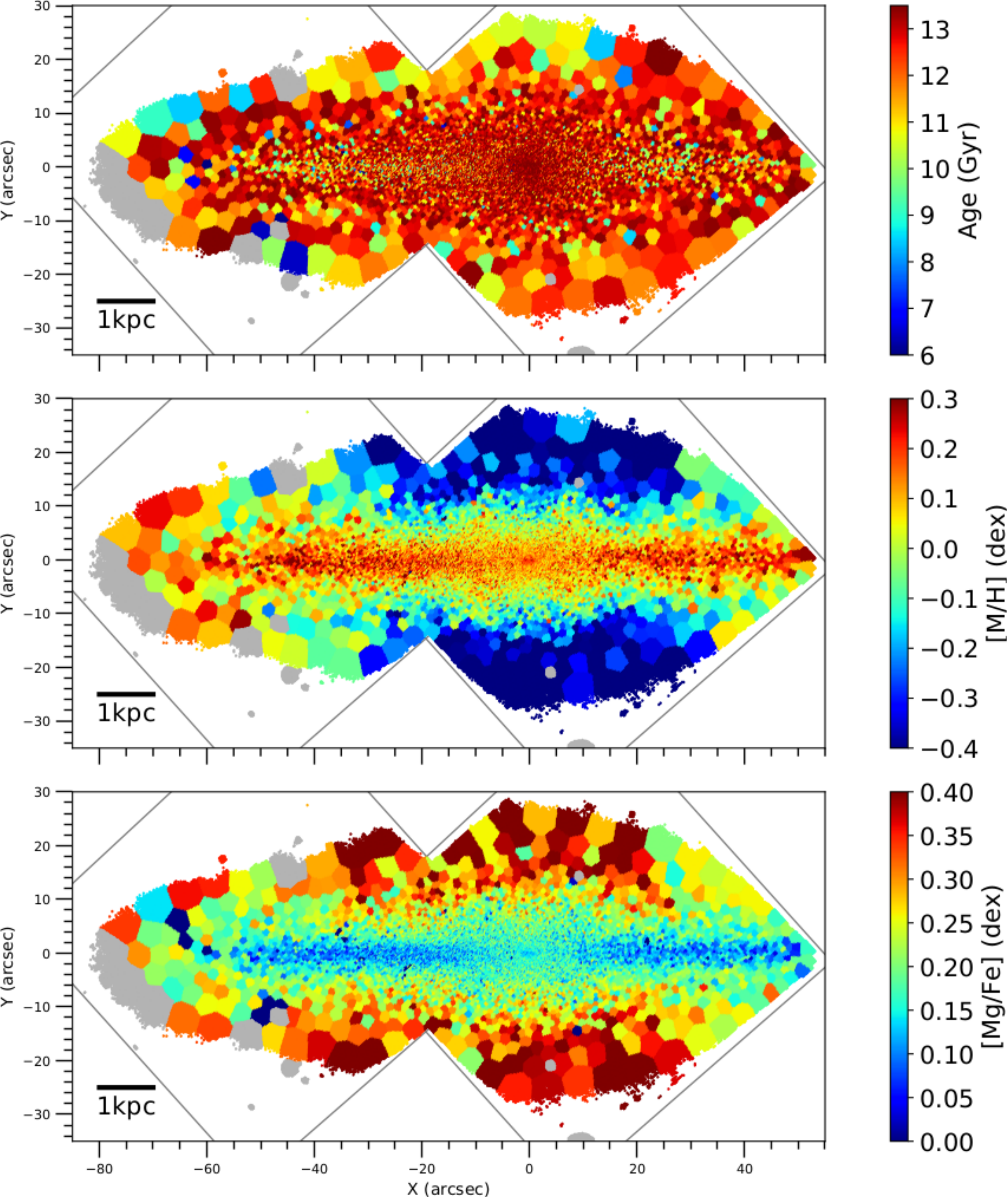}}
\caption{Maps of the stellar populations of FCC\,170. From top to bottom: mean age, total metallicity [M/H] and [Mg/Fe] abundance. The discarded bins and the position of the two MUSE pointings are plotted in grey. A scale bar on bottom-left
of each map indicates the correspondence with physical units.
}
\label{fig:pop}
\end{figure*}

In the velocity map (Fig.~\ref{fig:kin}, top panel), a clear spider diagram appears. 
We easily identify the fast-rotating thin disk and a slower-rotating thick disk at higher distance
from the midplane. It is also possible to see a faster-rotating nuclear disk.
Signs of cylindrical rotation are visible in the boxy-bulge region,
where the spider pattern is blurred and we see similar velocities at different distances from the midplane.
Nonetheless, the rotation is not perfectly cylindrical from our line of sight, probably due to the bar
orientation angle 
\citep[e.g.][]{Gonzalez2016,Molaeinezhad2016}.
The velocity dispersion shows a clear X-shaped pattern in the center, formed by a smaller X with higher $\sigma$ values and a bigger X with lower values. The small X presents a drop in the midplane, probably related to the presence of the nuclear disk.
The lowest velocity dispersions are found in the thin disk,
although low sigma values appear to extend also out from the midplane at larger radii (we show evidence of a thin disk flaring in \S~\ref{sec:pop_maps} and \S~\ref{sec:thick}).
This causes a radial gradient in the
thick-disk region (from $\sim$\,10\,arcsec onwards, see \S~\ref{sec:FCC170}), which
has globally lower $\sigma$ than the boxy bulge and higher $\sigma$ than the thin disk.\looseness-2

The skewness map describes the asymmetric deviations of the LOSVD from a Gaussian \citep{vanderMarel1993,Gerhard1993}. It clearly shows the different morphological structures of FCC\,170. An anticorrelation of $h_3$ with respect to the mean velocity is usually associated with disk-like components \citep[e.g.][]{Krajnovic2008,Guerou2016}. It indicates in Fig.~\ref{fig:kin} the positions where the thin and the nuclear disks are located. It also confirms that the box/peanut shape corresponds to a bar viewed edge on, by showing a correlation with the mean stellar velocity \citep[e.g.][]{Bureau2005}.
It indicates the region where this bar dominates the kinematics. 
In the kurtosis map in Fig.~\ref{fig:kin}, the different structures are barely visible. Positive values,
related to a LOSVD with a narrower symmetric profile than a pure Gaussian \citep{vanderMarel1993,Gerhard1993},
are predominant in the box/peanut region and negative values in the (thin) disk region. \looseness-2

\subsection{Mass-weighted stellar-population maps}\label{sec:pop_maps}
From the mass weights given by pPXF to the different SSP models, it is possible to reconstruct age, total metallicity [M/H] and abundance [Mg/Fe] of the stellar populations building the galaxy. 
Each SSP model corresponds to a combination of these three properties, therefore the weights indicate 
the importance of each combination in each individual bin.
By averaging the different populations making up each bin, with their respective weights,
we obtained the mean quantities plotted in Fig.~\ref{fig:pop}.
Uncertainties on the stellar-population parameters are given in Appendix~\ref{sec:unc}.

From the age map in the upper panel of Fig.~\ref{fig:pop}, we learn that FCC\,170 is overall old, 
with relatively small variations of age generally in the range between $\sim$\,8 and $\sim$\,14 Gyr. 
For this reason, it is difficult to clearly distinguish the different structures in this age map.
It is barely possible to recognize the oldest very central region and the youngest thin disk, 
while some thick disk bins look surprisingly younger. 
However, the transitions between the different structural components
are well visible in the mean metallicity map, in the middle panel of Fig.~\ref{fig:pop}.
A prominent X-shaped feature is traced by the metal-rich stars of the edge-on bar
(solar/supersolar metallicities), 
with the even more metal-rich nuclear disk in the center.
The most metal-rich region appears to be the thin disk, with no prominent radial gradient similarly to what has been found in the Milky-Way inner thin disk \citep{Fragkoudi2017b}.
The thin disk of FCC\,170 exhibits a metallicity flaring in the outer part and touches the thick disk region (10\,arcsec above the midplane) still with high metallicity. 
The thick disk displays rather subsolar metallicities in its inner part. 
In the outer part, a radial metallicity gradient is plainly
visible in the map. This could be related to the thin disk flaring, mixing more metal-rich populations, 
associated to the thin disk, with the most metal-poor, belonging to the thick disk.

In spite of having only two possible values of [Mg/Fe] in the SSP models, averaging them by their respective weights assigned by pPXF we obtained average [Mg/Fe] values in the full range between 0.0 and 0.4\,dex.
Several structures are also easily visible in the map of the mean abundance of [Mg/Fe]. In the bottom panel of Fig.~\ref{fig:pop}, the X-shape is 
suggested by the distribution of bins with [Mg/Fe] 
in the range between $\sim$\,0.05 and $\sim$\,0.10 dex.
The less $\alpha$-enhanced tiny 
nuclear disk is clearly visible. 
The thin disk appears as the least $\alpha$-enhanced component, with slightly supersolar values of [Mg/Fe], and a potential flaring 
connecting 
to the more [Mg/Fe]-enhanced outer thick disk.
The most $\alpha$-enhanced region is certainly the inner thick disk, with a difference of about 0.30\,dex 
with respect to the thin disk.

\subsection{Structural decomposition}\label{sec:morph}
The kinematic maps, especially the third-order Gauss-Hermite moment $h_3$, helped us to outline the regions where each structural component dominates. This is necessary to perform the stellar population analysis for each one of these components (see \S~\ref{sec:SFH} and \S~\ref{sec:chem}). 
It is possible to calculate the mean properties of the individual structures averaging on the bins dominated by it, or collapsing the spectra of these bins to obtain a single spectrum per component.
It is important to note that our comparison between components will be only qualitative, given that it is based on the line-of-sight properties. We cannot isolate the values for the individual components where they are superimposed, especially in the central region of the galaxy.
\begin{figure}
\scalebox{0.48}{
\includegraphics[scale=1]{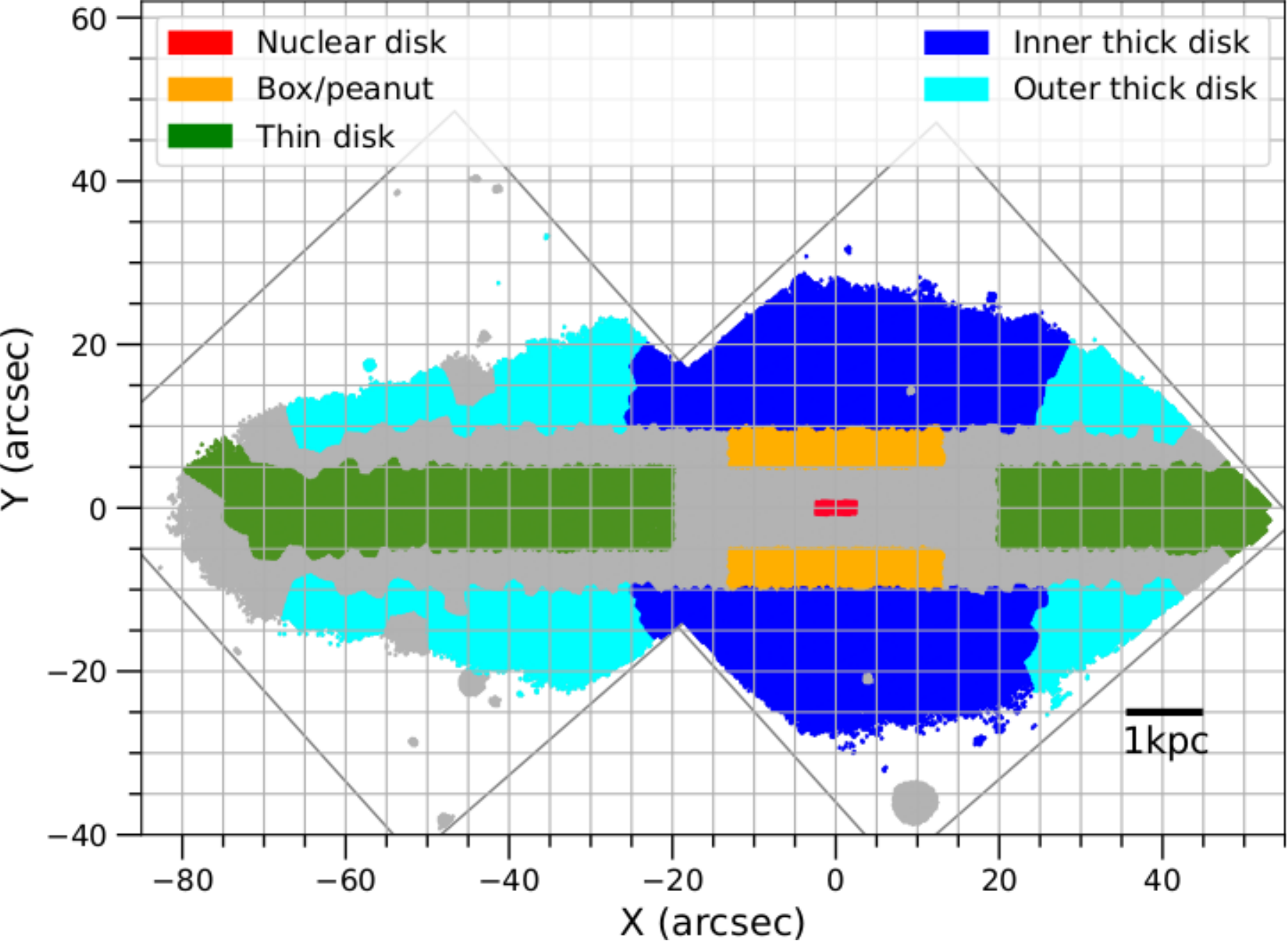}}
\caption{Map of the structural decomposition used in this work. Each color corresponds to a different component. The thick disk is divided into inner and outer regions. We plotted in grey all the bins not taken into account for the analysis of the single components, as well as the position of the two MUSE pointings. A scale bar on bottom-right indicates the correspondence with physical units.
}
\label{fig:morph}
\end{figure}

Following \citet{Comeron2018}, we consider the bins with $\abs{y}>10$\,arcsec dominated by the thick disk. 
We also divided the bins into an inner thick disk ($\abs{x}<25$\,arcsec) and an outer thick disk ($\abs{x}>25$\,arcsec),
since they appear to have slightly different chemical properties (see \S~\ref{sec:pop_maps}).
We define a thin-disk-dominated region between $\abs{y}<5$\,arcsec and $\abs{x}>20$\,arcsec, excluding a 
transition range between the thick and the thin disk ($5$\,arcsec\,$<\abs{y}<10$\,arcsec) and the central
region under the influence of the bar ($\abs{x}<20$\,arcsec). 
To study the boxy-bulge and the nuclear disk, we selected the regions where they dominate $h_3$. 
We selected the region with $\abs{y}<0.8$\,arcsec and $\abs{x}<2.5$\,arcsec for the nuclear disk and
the region with $5$\,arcsec\,$<\abs{y}<10$\,arcsec and $\abs{x}<13$\,arcsec for the box/peanut.
We observe these two components in the line of sight, overlapped to each other (in the case of the nuclear disk) and to the thin and the thick disk.
A quantitative analysis would be strongly influenced by other very bright components such as the thin disk. In particular, the properties obtained for the nuclear disk could be seriously affected by the ones of the box/peanut and the thin disk. Nevertheless, the data allows a qualitative comparison between components.

We show the bins selection for the individual structural components in Fig.~\ref{fig:morph}.
Martig et al. (in preparation) carried out an analysis on the galaxy NGC\,5746, very similar to this work.
They compared the decomposition method used here (based on kinematics) with a morphological decomposition based on surface brightness profiles. They confirm that the structural decomposition based on kinematics is the most restrictive, in the sense that it takes into account the regions least affected by the other components.\looseness-2

\subsection{Star formation history (SFH)} \label{sec:SFH}
We reconstructed the SFH of the individual bins of FCC\,170 summing the mass fraction of the different 
coeval SSP. 
We calculated the mass in each individual spatial bin, using the mass-to-light ratio in the $V$-band corresponding to each SSP model\footnoteref{note3}.
We first converted the $g$-band surface brightness in an image of the galaxy (from the Fornax Deep Survey, FDS, \citealt{Iodice2016}) to the $V$-band in the Vega system.
Then, we compared the surface brightness in our MUSE spectra to the $V$-band surface brightness in the image, in an annulus between 5 and 10\,arcsec, to obtain an offset 
between them.
Taking into account this offset and the galaxy distance (\S~\ref{sec:FCC170}), we estimated the $V$-band absolute surface brightness in each spaxel of our MUSE data.
The mean $M$/$L$ per spaxel was calculated as the weighted average of the $M$/$L$ corresponding to the models assigned to that spaxel by the pPXF best fit.
We used Eq.~1 in \citet{Cebrian2014} to compute the stellar mass per spaxel from the $M$/$L$ and absolute surface brightness (both in the $V$-band).
This method gave us a total stellar mass of about $5.8\times10^{10}$\,M$_{\odot}$ in the region of FCC\,170 covered by our Voronoi binning, a lower limit for the galaxy stellar mass.

\begin{figure*}
\centering
\resizebox{0.9\textwidth}{!}
{
\includegraphics[scale=0.46]{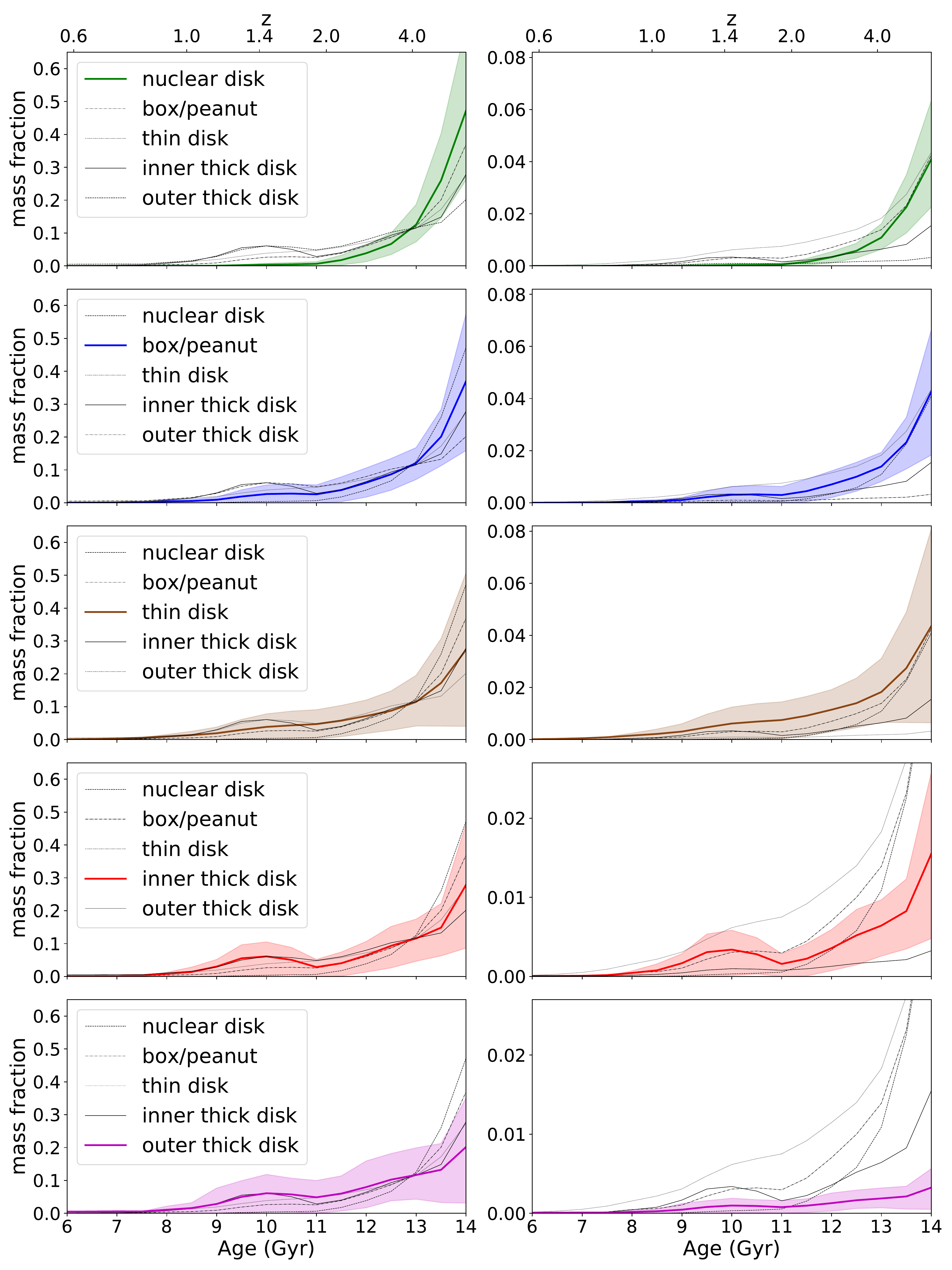}
}
\caption{Weighted average SFH of the different structural components of FCC\,170, as defined in Fig.~\ref{fig:morph}.
The average mass fraction is displayed on the vertical axis, weighted by the mass in each bin.
This mass fraction is normalized to the total mass of the component, in the left column, and to the total mass of the galaxy, in the right column.
In each panel, only one component is represented in a specific color (\textit{green} for the nuclear disk, \textit{blue} for the box/peanut, \textit{brown} for the thin disk, and \textit{red} and \textit{purple} respectively for the inner and the outer thick disk). 
For this component, 1$\sigma$ uncertainties, calculated as 16\,\% and 84\,\% percentiles of the spatial-bin distribution, are represented as shades of the same color. The SFH of the rest of components are plotted in black, with different line styles, in each panel.
As reference, the redshift scale on top of each column indicates the epoch when stars of a given age were born. The correspondence between redshift and age was approximated with Eq.~2 in \citet{Carmeli2006}.
}
\label{fig:sfh_4c}
\end{figure*}

Following the spatial-bin decomposition in \S~\ref{sec:morph} we averaged, on the different bins 
of a specific individual structural component,
the mass fraction corresponding to the same age.
We weighted according to the different masses in the individual spatial bins (for the specific age) when averaging.
This method results in the SFHs of the different components shown in Fig.~\ref{fig:sfh_4c}.
The left column, where the mass fraction was normalized to the stellar mass of the corresponding structural component, is more recommended to compare the different populations in the same component. In the right column, where the mass fraction was normalized to the galaxy total mass, we can compare the different components with each other.
The thin disk is clearly the most massive for all ages.
This SFH is non-cumulative in the sense that it does not show the properties of the global  galaxy for a given time, but only the properties of specific stellar populations (with the age indicated in the plots).
In each panel, only the average SFH of one component is plotted with a specific color and with upper and lower limits of the same color.
These indicate the 1$\sigma$ dispersion between the individual spatial bins of each component. They were calculated as the 
16\,\% and 84\,\% percentiles of the mass fraction distribution over the different bins.
The average SFHs of the other structural components were plotted in each panel in black with different line styles, for reference.
We compared these results with those obtained from collapsing all the bins of a component into a single spectrum.
We found similar star formation histories with the two methods.

The SFH of the four components peak around 14\,Gyr, confirming the predominance of the 
very old populations in the mean age map 
in Fig.~\ref{fig:pop}.
The similarity in the shapes of the curves in Fig.~\ref{fig:sfh_4c} is probably favored by the line-of-sight overlapping, in some regions (especially the central region), of stars belonging to different components (see also \S~\ref{sec:morph}), 
and by our age resolution (see Appendices~\ref{sec:unc} and \ref{sec:tests}).
However, the structural components differ in the sharpness of this peak 
and the extent of the SFH.
In the nuclear-disk region (green curve in the first row from top of Fig.~\ref{fig:sfh_4c}), for instance, stars formed in a very early epoch and star formation roughly stopped around 11\,Gyr.
We cannot rule out the presence of an older central component, i.e. a classical bulge, difficult to see in our maps because of its relatively reduced size, mass and brightness. It would be contaminating the SFH of the nuclear disk.
The box/peanut
(blue curve in the second row from top) has also a very pronounced 
peak around 14\,Gyr, but a slightly more extended SFH including a very 
smooth peak around 10\,Gyr.
This peak is much more pronounced in the thick disk, where
the star formation had almost halted around 11\,Gyr,
when this
younger population appeared.
This is present in both the red and purple curves in the two bottom rows of Fig.~\ref{fig:sfh_4c}, 
corresponding to the inner and outer thick disk, respectively.
The red and purple shades show that this bump is present in all the spatial distribution.
The thin-disk SFH appears to be more continuous (brown curve in the third panel from top), extending down to $\sim$\,7\,Gyr.
The fact that the 10\,Gyr peak is more pronounced in the inner than in the outer thick disk is probably related to the thin-disk flaring.
The outer-thick-disk SFH is compatible with a composition of 
inner-thick-disk-type populations 
with thin-disk populations.

\subsection{Chemical abundances} \label{sec:chem}
\begin{figure}
\resizebox{.485\textwidth}{!}
{
\includegraphics[scale=0.35]{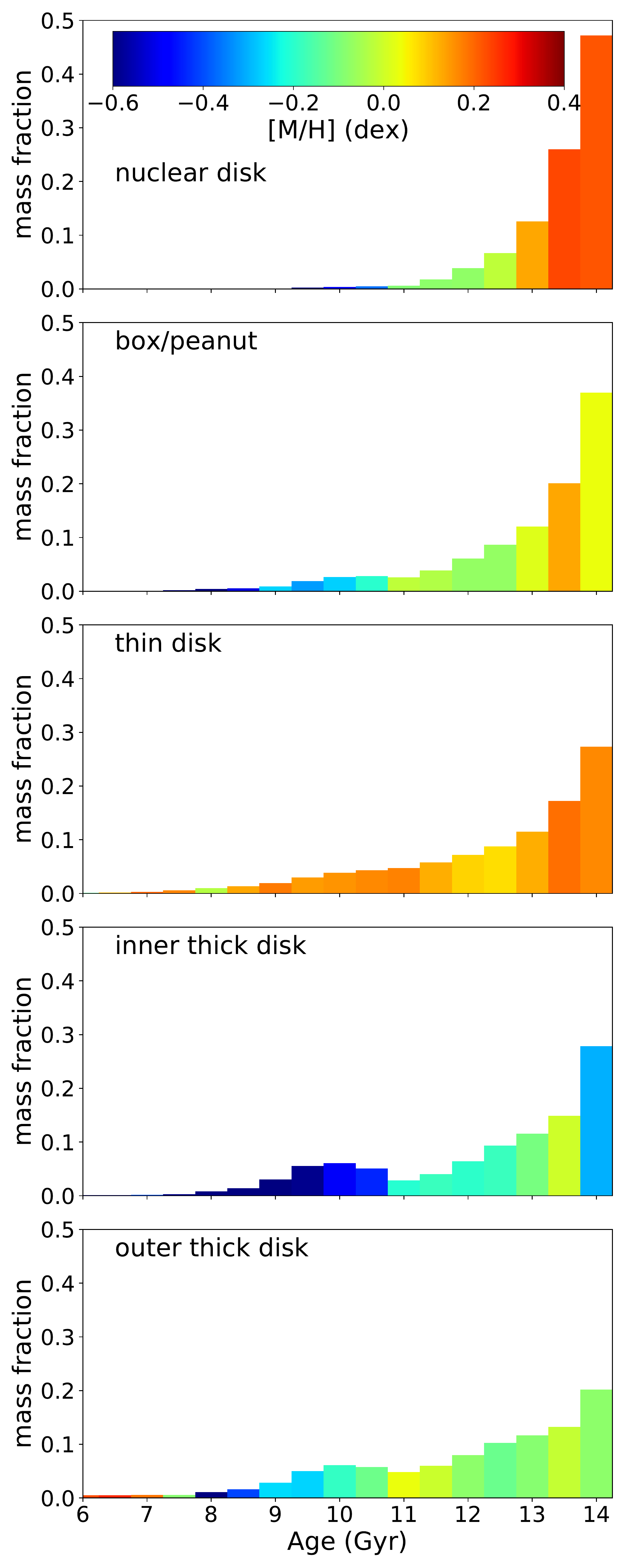}}
\caption{SFH of the different structural components making up FCC\,170. 
From top to bottom: the nuclear disk, the box/peanut, the thin disk, the inner and the outer thick disk.
We have color coded the histogram age bins according to the weighted average total metallicity ([M/H]) of the specific age bin.
The average mass fraction is displayed on the vertical axis, weighted by the mass in each bin and normalized to the mass of the component.
}
\label{fig:agemet_hist_comp}
\end{figure}

The chemical composition of the
populations appearing in the SFH
can help us to understand if they  
came from different formation processes or if they were simply part of a continuous build-up of the galaxy.
For each structural component, 
in Fig~\ref{fig:agemet_hist_comp} we have plotted histograms of the SFHs in the left column of Fig.~\ref{fig:sfh_4c}, with the mass fraction normalized to the component mass.
This time we have color coded the different age bins according to the weighted average of the total metallicity 
(by weighting on the different SSP models according to their weights and on the spatial bins according to their mass). 
Most stars formed from an evolved interstellar medium, since in general we measure
solar or supersolar metallicities.
The most metal-rich structures are the nuclear and the thin disks (first and third panels from top), while the thick disk clearly displays subsolar metallicities (last two panels from top).
The most metal-poor stars (reaching [M/H]$\lesssim -0.5$\,dex) form the younger peak at $\sim$\,10\,Gyr in the thick-disk SFH.
These stars formed later but from a less chemically evolved gas. 
This difference in metallicity might be the hint of a different formation history for this younger population, 
compared to the older and less metal-poor rest of the thick disk.
Comparing the inner and the outer thick disks, 
the $\sim$\,10\,Gyr-old population is clearly more metal-poor, on average, in the former.
The metallicity in the outer 
region displays one more time the mixing of the metal-poor populations with the metal-rich stars in the flaring thin disk.
The smooth younger peak visible in the second panel from top, in Fig.~\ref{fig:agemet_hist_comp}, is also more metal-poor than the rest of the box/peanut, 
suggesting that the same thick disk younger population might be present also here, although to a minor extent.

\begin{figure*}
\centering
\resizebox{0.98\textwidth}{!}
{
\includegraphics[scale=0.9,width=\textwidth]{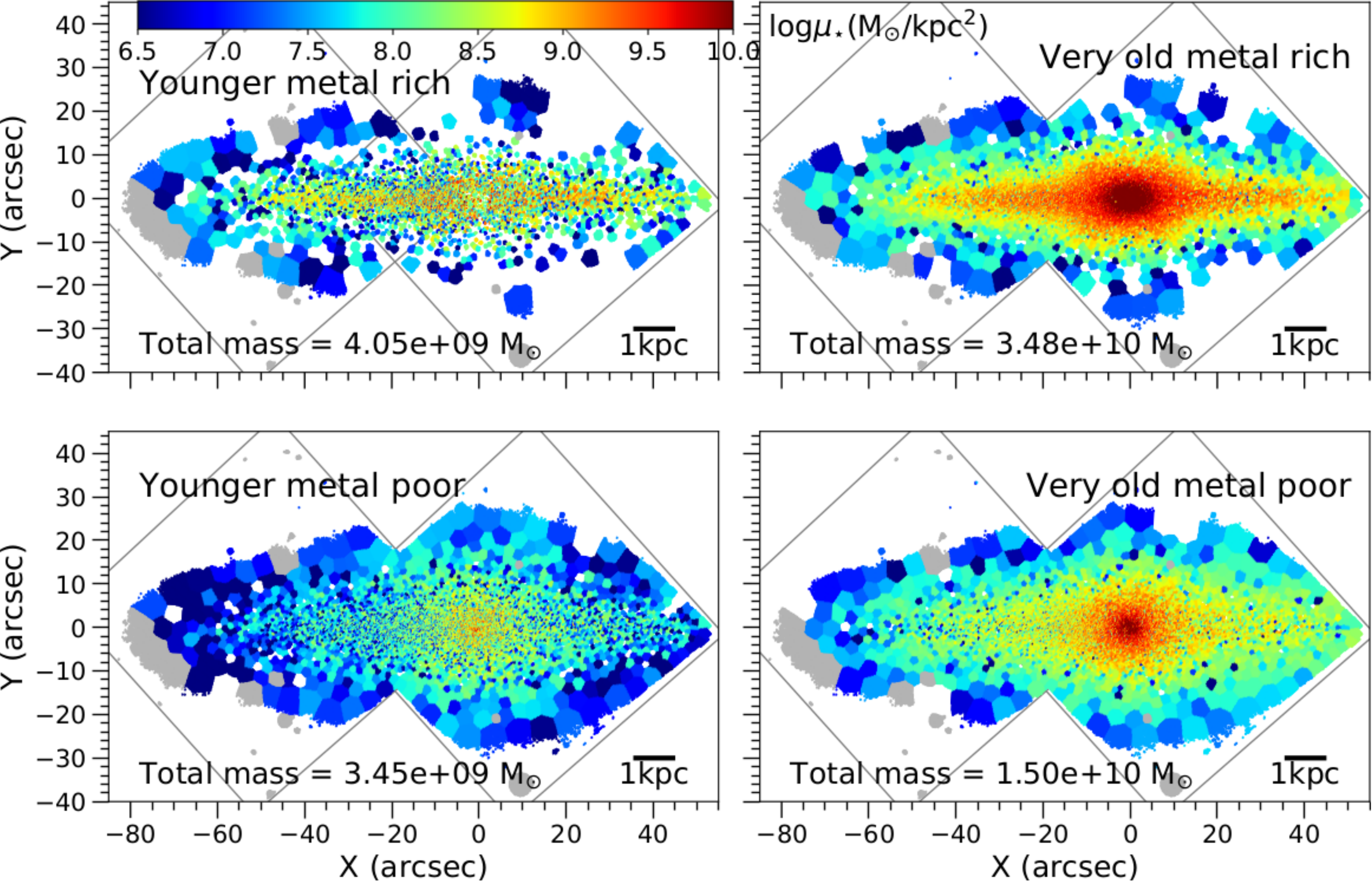}}
\caption{Maps of some combinations of age and metallicity for the stellar populations in FCC\,170.
\textit{Top}: most metal-rich populations ([M/H]$\ge 0.06$\,dex).
\textit{Bottom}: most metal-poor populations ([M/H]$\le -0.25$).
\textit{Left}: little younger populations (ages $\le$\,11.5\,Gyr).
\textit{Right}: oldest populations (ages $\ge$\,11.5\,Gyr).
The color scale shows the mass density corresponding to the populations in the specific age-metallicity
bin.
The total mass in the age-metallicity bin is indicated on bottom-left of each map.
The position of the two MUSE pointings is plotted in grey. A scale bar on bottom-right
of each map indicates the correspondence with physical units.
}
\label{fig:agemet_bins}
\end{figure*}

\begin{figure*}
\centering
\resizebox{0.8\textwidth}{!}
{
\includegraphics[scale=1,width=\textwidth]{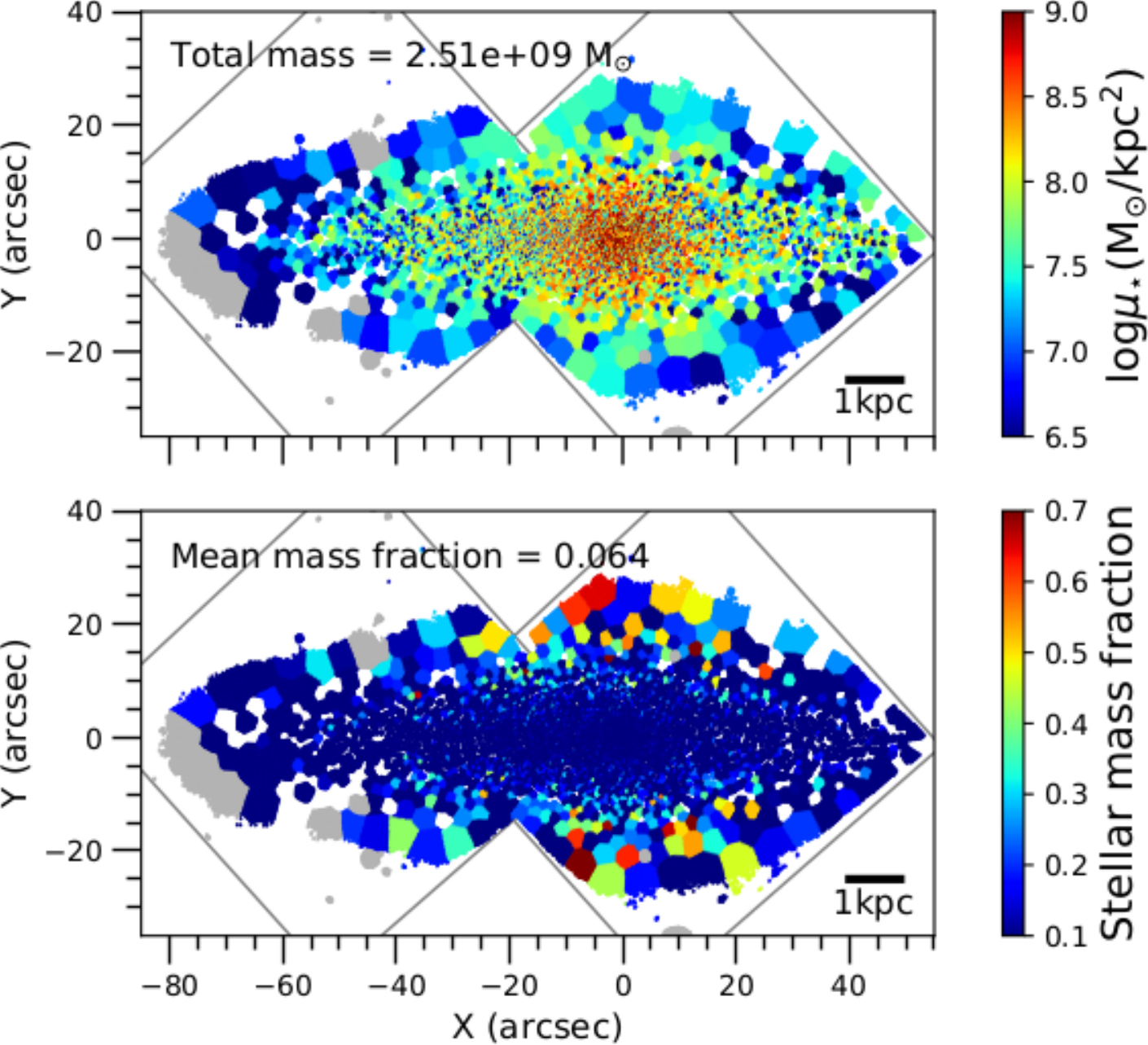}}
\caption{Distribution of the stellar population with ages between 9 and 11\,Gyr, metallicity between -0.96 and -0.06\,dex and [Mg/Fe]$\sim$0.4\,dex.
The color scale shows the stellar mass density in the top panel and the stellar mass fraction in the bottom panel.
The total mass or the mean mass fraction in this specific population are indicated on bottom-left of respectively the top and the bottom maps.
In each panel, the position of the two MUSE pointings is plotted in grey and a scale bar on bottom-right indicates the correspondence with physical units.
}
\label{fig:accreted}
\end{figure*}

Dividing the age and metallicity ranges into different bins, it is possible to plot the spatial distribution of the different stellar populations.
In Fig.~\ref{fig:agemet_bins}, we have mapped the mass density of the different stellar populations according to the combination of their age and metallicity.
The top-left panel shows that younger and more metal-rich stars are not contained in all spatial bins and are mainly in the thin disk and the box/peanut.
Most metal-rich stars are very old (i.e. older than 11.5\,Gyr, see top-right panel), contributing about half the total mass. They form almost entirely the box/peanut and their presence is very strong in the inner thin disk too. They are also present in the thick disk in a minor extent.
Although most metal-poor stars dominate the thick disk in mass fraction, their mass is concentrated in the very central region, suggesting a spherical distribution (bottom panels in Fig.~\ref{fig:agemet_bins}). This could be a small classical bulge that, because of its more reduced size than the bar, is not visible in the other maps. In contrast, in the same region, metal-rich stars followed the shape of the box/peanut (see top panels).
The population corresponding to the secondary peak in the thick-disk SFH plot is included in the bottom-left panel of Fig.~\ref{fig:agemet_bins}, but has been also isolated in Fig.~\ref{fig:accreted}.
Here we see that it is mainly concentrated in the central region and the inner thick disk but present everywhere (top panel). It contributes 
$\sim 2.5\times 10^9$\,M$_{\odot}$ in the region under study in this work, approximately the 10\% of the galaxy total stellar mass.
The bottom panel shows that this population dominates several less dense bins of the inner thick disk, while in the rest of the galaxy it contributes only about the 6\% of the bin mass, on average.

\begin{figure}
\resizebox{.485\textwidth}{!}
{
\includegraphics[scale=0.39]{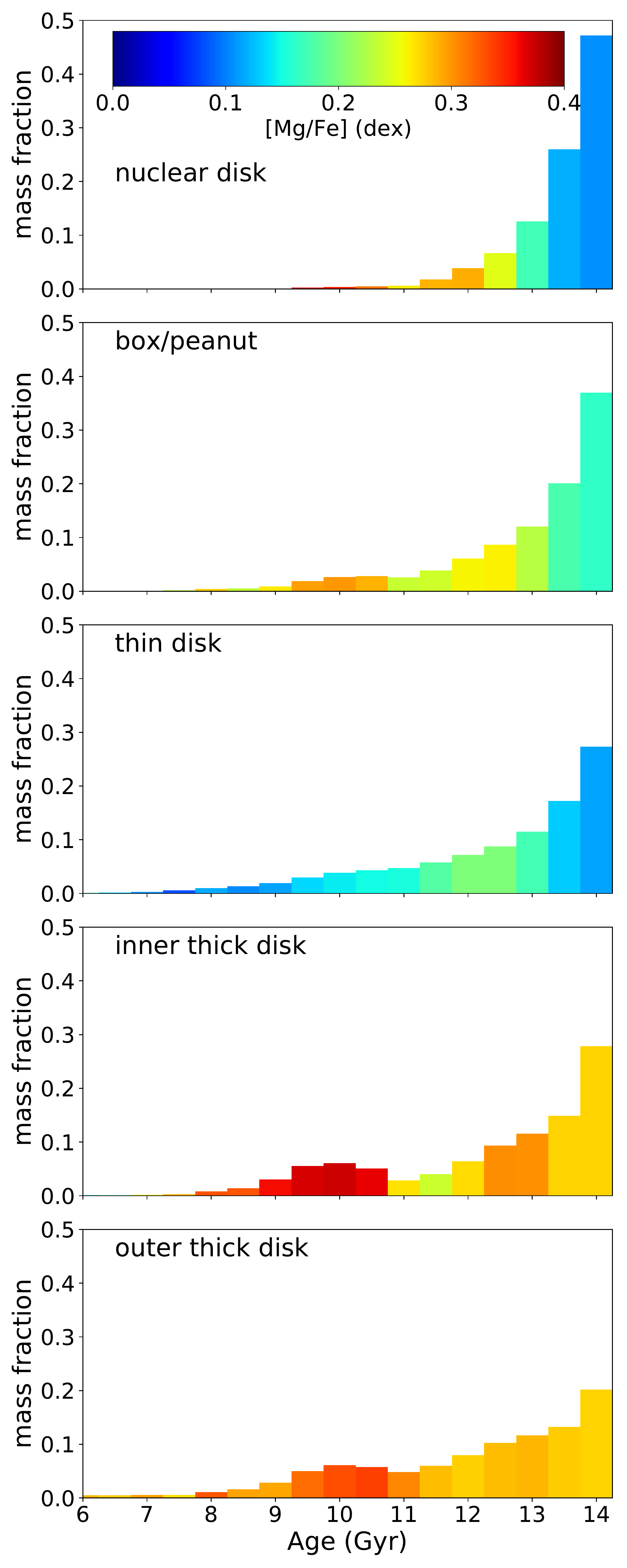}}
\caption{SFH of the different structural components of FCC\,170. 
From top to bottom: the nuclear disk, the box/peanut, the thin disk, the inner and the outer thick disk.
We have color coded the histogram age bins according to the weighted average [Mg/Fe] of the specific age bin.
The average mass fraction is displayed on the vertical axis, weighted by the mass in each bin and normalized to the mass of the component. 
}
\label{fig:agealpha_comp}
\end{figure}

The [Mg/Fe] abundance can add some insights to our analysis. In fact, the abundance ratio between an
$\alpha$-element and iron can be used as a "star formation clock"
\citep[e.g.][]{Peletier2013}. 
It gives valuable information about the formation timescale of a galaxy or part of it. This ratio would be higher if stars were formed rapidly, since $\alpha$-elements are 
mostly produced by supernovae type II, than if they were formed slowly, giving time to supernovae type Ia 
to produce Fe-peak elements.
In Fig.~\ref{fig:agealpha_comp} we show the SFH histograms of the different components, this time color coding the different age bins according to the corresponding mean value of [Mg/Fe]. 
These plots indicate that the $\alpha$-enhancement increases towards younger populations except in the thin disk, less enhanced in [Mg/Fe] for all ages. The thick disk is globally the most $\alpha$-enhanced component. In particular, the distinction between the oldest populations and the bump in the SFH (peaking at $\sim$\,10\,Gyr)
is emphasized by a jump to a higher [Mg/Fe] abundance, both in the outer and the inner thick disks.
This means that these younger stars were formed not only from a less evolved interstellar medium, but also with a different formation timescale, compared to the rest of the thick disk. 
Our results point to different formation scenarios for this younger population with respect to the dominant very old ones.

\section{Discussion}
\label{sec:discussion}
\subsection{Comparison with previous works on FCC\,170} \label{sec:comparison}
Our work, making use of integral-field data, shows 
a two-dimensional view of FCC\,170 stellar
kinematics and populations.
Previous studies used only long-slit observations and are not straightforward to compare to.
In their kinematic profiles from long-slit observations,
several authors \citep{Chung2004,Bedregal2006,Spolaor2010a,Koleva2011,Williams2011} already identified  some signatures corresponding to the structures in our kinematics maps, such as humps in the rotation curve or $h_3$ correlations and anticorrelations with the mean velocity. 
Therefore, our bi-dimensional results confirm the presence of the structures associated to these
kinematic signatures: the nuclear disk, the 
bar seen as a boxy-bulge, the thin disk. 
The compatibility of the positions dominated by these components, as determined from our maps and from the rotation curves in previous works, allowed us to define them spatially. 
\citet{Williams2011}, in particular, measured the stellar kinematics of FCC\,170 with four slits at different distances from the midplane. 
However they could not measure many points at disk-dominated radii. 
They found signs of cylindrical rotation, which is visible in our mean velocity map in Fig.~\ref{fig:kin}. 
Their velocity dispersions at the different slit positions also agree very well with our maps, with a central peak of $\sim$\,160\,km\,s$^{-1}$ and values between 80 and 100\,km\,s$^{-1}$ in the thick disk region.
Our maximum rotation velocity and velocity dispersion are also in very good agreement with the rotation curve and $\sigma$ profile presented in \citet{Chung2004}.

When it comes to the stellar populations, we have partial agreement with some of the studies in the literature.
\citet{Spolaor2010b} found FCC\,170 uniformly old along the major axis, with (luminosity-weighted) ages compatible with our
Fig.~\ref{fig:pop}. 
Regarding metallicity, they found a radial profile characterized by a steep gradient, spanning from supersolar to subsolar (luminosity-weighted) values. 
In the same radial extent, the midplane is overall metal rich according to our analysis, instead.
Analysing the same data of \citet{Spolaor2010b} with different methods, \citet{Koleva2011} found a more uniform metallicity along the major axis, at about solar [Fe/H], except for the outskirts (from about 60\,arcsec on) where it dropped to subsolar [Fe/H] values and where also our [M/H] decreases.
Our age results are in a good agreement with \citet{Williams2011}, who observed overall old populations in the region covered by their four slits, including the thin disk and up to 12\,arcsec
above the midplane.
They detected a negative vertical gradient from solar to subsolar metallicities, and a positive vertical [$\alpha$/Fe] gradient, starting from a solar abundance in the midplane up to
supersolar.
Also, \citet{Johnston2012} found ages in the same range as we did, older than 10\,Gyr.
Our estimate of the lower limit for stellar mass (\S~\ref{sec:chem}) is in agreement with the total stellar mass estimated by \citet{Vaddi2016}.

\subsection{Formation and evolution scenarios for FCC170} \label{sec:formation}
The absence of structures in the age map of FCC\,170 is striking.
The spatial correspondence of the different structures in kinematics, metallicity and [Mg/Fe] abundance is not present in the age map. FCC\,170 looks overall old.
On the one hand, this is surprising given the results in the literature showing, by and large, a younger thin disk than the thick disk (e.g., in the Milky Way, \citealt{Gilmore1983,Prochaska2000,Cheng2012,Haywood2013} and in external galaxies \citealt{Yoachim2008b,Comeron2015,Guerou2016,Kasparova2016}).
On the other hand, 
massive
galaxies are expected to have been more efficient in forming stars, to have formed them quickly at high redshift, and therefore to be overall old and more $\alpha$-enhanced in their thick disks \citep[e.g.][]{Elmegreen2017}.
Furthermore, they tend to be globally more metal-rich \citep[e.g.][]{Spolaor2010b}.
FCC\,170 can be considered a
massive galaxy in this context.

Other studies also found similar ages in both disks.
No important age difference was found in the S0a NGC\,4111 by \citet{Kasparova2016}. Nevertheless, this galaxy was much younger than FCC\,170 ($\sim$\,5\,Gyr old)
and the chemical composition appeared very similar in both disks.
Another S0, ESO\,243-49, located in the Abell\,2877 cluster, did not display important differences in age between the thin and the thick disk.
It was analysed by \citet{Comeron2016} using MUSE data.
There are some similarities between their maps and the ones we show in this work.
ESO\,243-49 appeared to be overall old like FCC\,170, with no younger thin disk. 
Differences in metallicity were found between the thin and thick disk regions in ESO\,243-49, 
although less prominent than in FCC\,170.
In both galaxies these differences were not reflected in age.
Both galaxies are members of a cluster, being the environment potentially important for their evolution (see discussion in \S~\ref{sec:env}).
\citet{Comeron2016} adopted a "monolithic collapse" scenario \citep[e.g.][]{Eggen1962} to explain the properties of ESO\,243-49. 
The origin we propose for FCC\,170 differs from a simple monolithic collapse, but we do agree with a relatively fast formation at high redshift of
both disks, approximately equally old, which stopped forming stars very early.
In spite of this, the metal-rich thin disk (as well as the bar and the nuclear disk), 
%
had probably little more time to let the gas evolve from subsolar to supersolar metallicities.
Moreover, this little time let supernovae type-Ia increase their contribution compared to supernovae type-II, resulting in a lower $\alpha$-enhancement with respect to the thick disk \citep[e.g.][]{Brook2004,Yoachim2008b}.

Only if star formation in both disks occurred in a relatively short timescale, the age differences between the structural components would have become unrecoverable with our method (Fig.~\ref{fig:pop}).
The drop in the sensitivity of the H$\beta$ line depth to stellar ages from about 4\,Gyr on, in our models \citep{Vazdekis2010,Vazdekis2015}, together with the age-metallicity degeneracy, makes it arduous to distinguish between different values in the age range of FCC\,170 (see also Appendix~\ref{sec:tests}).
This could also explain why, in Fig.~\ref{fig:agemet_hist_comp} and \ref{fig:agealpha_comp}, nuclear disk, box/peanut and thin disk seem to be born already with high metallicities and low [Mg/Fe], while the Milky Way disk reached supersolar metallicities only after $4-5$\,Gyr \citep{Haywood2013,Snaith2014}.
According to the closed-box model, a very high star-formation efficiency would allow an initial almost instantaneous increase of metallicity, even up to supersolar values \citep{Vazdekis1996}.
This initial very high star-formation efficiency likely drove the early evolution of FCC\,170, since it formed a large stellar mass in a few Gyr.
In this case, the initial high metallicity would be expected to be associated to high values of [Mg/Fe], but in this metallicity regime our method could be less sensitive to variations in [Mg/Fe], so that we do not detect them \citep{Vazdekis2015}.

The chemical evolution shown in Fig.~\ref{fig:agemet_hist_comp} and \ref{fig:agealpha_comp} is somewhat counter-intuitive.
After a fast increase in the first Gyr, mean metallicity of recently born stars decreases with time (see e.g. the nuclear disk), contrary to what is usually predicted in a closed box.
In addition, we would expect Fe-peak elements to increase with respect to $\alpha$-elements for later generations of stars. This is not shown in the full age range by our plots, where [Mg/Fe] also increases with time in some epochs depending on the structural component.
This behavior might be either real or not (or partially real) due to a combination of the following reasons.
On the one hand, the oscillations of [M/H] and [Mg/Fe] with age might be due to technical issues, such as the drop of H${\beta}$ sensitivity in the age range of FCC\,170.
Also line-of-sight effects could be playing a role, in particular in the analysis of nuclear disk and box/peanut.
Stars of overlapped structural components, which we are unable to separate, might be dominating specific age bins.

On the other hand we cannot discard the reality of this chemical evolution, counter-intuitive from the point of view of a closed box.
It could be due to the accretion of gas less enriched than the host galaxy or accretion of stars with different compositions and ages.
They could have happened in several events during the life of FCC\,170 or in the single event that brought new populations to the thick disk, as it will be discussed later in this section.
Regularization could be responsible for diluting the accreted populations in a wider age range.
In the accretion case, the closed-box model could obviously fail.
However, a very high star-formation efficiency would allow, even in a closed box, a mild decrease in metallicity related to a hypothetical lack of available gas, after the initial very steep increase of metallicity \citep{Vazdekis1996}.
We advice the reader not to interpret our plots for a quantitative analysis based on the absolute values. Apart from what explained in the last paragraphs, these might be also affected by the limited age, [M/H] and [Mg/Fe] ranges of the specific SSP models. The fact that the results are mass-weighted should be additionally taken into account.
A qualitative analysis based on the mean values per bin should be safe given the spatial coherence between age, metallicity and [Mg/Fe].

\subsubsection{FCC\,170: in the core of the Fornax cluster} \label{sec:env}
FCC\,170 is located in the densest region of the Fornax cluster, where several signs of gravitational interactions have been found. 
\citet{Iodice2016} found a faint stellar "bridge" between NGC\,1399, the cD galaxy in the center of the cluster, and NGC\,1387. This indicates an interaction between the two galaxies, probably stronger in an earlier epoch.
In fact, \citet{Iodice2017a} detected intra-cluster light in a quite extended region of the cluster core, including FCC\,170.
In addition, this light was shown to be the counterpart of previously detected over-densities of blue globular clusters \citep{Bassino2006,dAbrusco2016}, indicators of a tidal stream.
The morphology, the location and the faintness of this ICL were interpreted by \citet{Iodice2017a} as signs of tidal stripping from the galaxies in this dense region, which would have passed close to the central cD in an earlier epoch.
The velocity difference of FCC\,170 with respect to NGC\,1399 is compatible with tidal interactions with the cluster potential \citep{Iodice2017a}, which could have strongly affected its evolution. 

Gravitational interactions in galaxy clusters are known to both increase and quench star formation \citep[e.g.][]{Fujita1999,Quilis2000,Kronberger2008,Cohen2014,Hwang2018}. 
"Ram pressure" can trigger or enhance efficient star formation in molecular clouds, during the interaction of a galaxy with the hot intra-cluster medium \citep{Bekki2003}. 
In particular, during the galaxy first infall into the cluster, the fast increase in the external pressure can trigger a starburst \citep[e.g.][]{Evrard1991}.
In addition, \citet{Byrd1990} and \citet{Hwang2018} showed that tidal and hydrodynamic interactions with the neighbor galaxies are more than sufficient to enhance star formation by induced collisions and compression of disk gas clouds.

On the other hand, several processes have been proved to strip the gas from galaxies in dense environments, as confirmed by the detection of molecular gas in tidal tails \citep[e.g.][]{Kenney1999,Braine2000,Verdugo2015}.
\citet{Quilis2000} and \citet{Abadi1999}, among other authors, claimed that ram pressure and viscous stripping are very effective at removing the disk gas in a relatively short timescale. 
In a longer timescale, "harassment" is another process that can strip material from galaxies, forming tidal tails \citep[e.g.][]{Moore1996}.
Finally, the stripping of the hot halo gas causes "strangulation", i.e. the slow decline in the infall of fresh gas into the galaxy disk, gradually truncating the star formation \citep[e.g.][]{Balogh1999,Balogh2000a,Balogh2000b,Bekki2009,Bekki2010}.

\citet{Fujita1999} found that the star formation rate can even double in the initial approach of the galaxy to the center of a dense cluster, but then drops fast due to the gas depletion from stripping mechanisms, when the galaxy has reached the cluster core.
However, the intense star formation and the posterior quenching happened in FCC\,170 rather early in the life of the Universe, when probably the Fornax cluster was not formed yet as we observe it today.
According to \citet{Chiang2013}, a Fornax-type cluster would have reached half of the current mass at redshift $z\sim 0.6$, while at $z\sim 2$ it would have had only the 10\%.
Galaxy clusters have been found up to $z\sim 2$, but protoclusters can be already identified at $z\sim 6$ in cosmological simulations \citep{Overzier2016}.
The progenitors of nowadays's groups and clusters had a very intense star formation, shaping the massive galaxies that we observe today. The more massive the halo, the higher the star formation rate.

As pointed out by \citet{Iodice2018}, FCC\,170 has a receding velocity similar to other three galaxies in the same region of the cluster (FCC\,167, FCC\,182 and FCC\,190, see e.g. \citealt{Ferguson1989a}), while $\sim$300\,km s$^{-1}$ larger than the central galaxy FCC\,213.
This suggests that these four galaxies, aligned in the north-south direction, belonged to a primordial (sub-)group, before falling into Fornax.
That could be the environment favoring the 
initial intense star formation in FCC\,170,
giving rise to a relatively large stellar mass fast and early.
Its gas could have been (partially) exhausted, and the remaining amount (if any) could have been stripped out later by interactions of different nature.
The star formation was quenched very early.
This "pre-processing" before entering the current-day cluster \citep{Haines2012,Haines2015} would have given already the observed appearance to FCC\,170, with
roughly no gas nor dust, nor young populations, and could explain the shape of its SFH (as shown in \S~\ref{sec:SFH}).

\subsubsection{Signs of accretion in the thick disk} \label{sec:thick}
In FCC\,170, the thick disk is clearly made up by the contributions of different populations, as shown in 
\S~\ref{sec:SFH} and \ref{sec:chem}. 
The thick-disk main population was probably born already thick.
It is old and its composition is different from the rest of the galaxy.
Its $\alpha$-enhancement 
indicates that it must have formed relatively faster.
However, the presence of the second peak in the SFH (Fig.~\ref{fig:sfh_4c})
calls for a different origin for this younger population.
Its chemistry is very different not only 
from the rest of the galaxy but also from the older population in the thick disk (Figs.~\ref{fig:agemet_hist_comp} and \ref{fig:agealpha_comp}).
This population formed later, from a less enriched gas and in an even shorter timescale than the oldest population.
An external origin could explain these properties.
Stars would have been formed, in a satellite with its own chemistry,
about 10\,Gyr ago.
The satellite would have had a stellar mass of at least $2.5\times 10^9$\,M$_{\odot}$, according to Fig.~\ref{fig:accreted}, taking into account that this lower limit was obtained only from the region covered by our data and the satellite could have lost material at larger distances from FCC\,170 during the early phase of accretion.
Its accretion could have preceded the onset of type-Ia supernovae in its stellar population, leaving its stars with the observed high values of [Mg/Fe].
As suggested by \citet{Qu2011}, its stars could have been initially distributed with an increasing fraction with height.
The bottom panel of Fig.~\ref{fig:accreted} indicates that they dominate several bins of the inner thick disk.
Nonetheless, the top panel shows that these stars probably had the time to be redistributed overall the galaxy and to migrate mostly to the central region.

FCC\,170's formation could have much in common with the "two-phase scenario".
This scenario was initially proposed for early-type galaxies by \citet{Oser2010}, based on their simulations.
It includes a first \textit{in-situ} phase at high redshift ($z\gtrsim 2 - 3$) followed by a second \textit{ex-situ} phase at intermediate or low redshift ($z\lesssim 2 - 3$).
The first phase would have been dominated by an \textit{in-situ} star formation related to cold gas inflows.
This phase is
characterized by an early star formation peak which is relatively extended in time.
It could explain the main very old stellar populations in all the components of FCC\,170.
The second phase would have been characterized by the accretion of stars from satellites and
could explain the secondary peaks in the SFH of the thick disk and box/peanut of FCC\,170.

\citet{Yoachim2005} invoked satellite 
accretion to explain the thick-disk kinematics
of the two galaxies in their sample.
In addition, the variety of the kinematics observed by
\citet{Yoachim2008a} confirmed the compatibility with an external origin.
In one galaxy in the sample of 
\citet{Yoachim2008b}, 
stars appeared to be more $\alpha$-enhanced at larger radii, also
interpreted as a later accretion.
\begin{figure*}[h!]
\centering
\resizebox{0.9\textwidth}{!}
{
\includegraphics[scale=0.36]{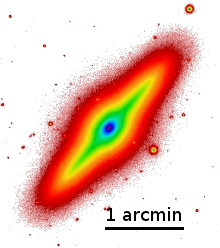}
\hspace{1mm}
\includegraphics[scale=0.36]{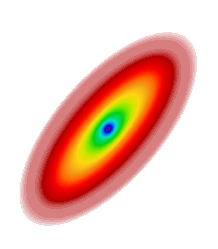}
\hspace{1mm}
\includegraphics[scale=0.36]{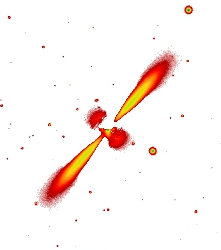}
}
\caption{Isophotal analysis for FCC\,170 in the \textit{r}-band from \citet{Iodice2018} (Fig.\,14). From left to right: data from the FDS \citep{Iodice2016}, bi-dimensional model and residual image. The latter was obtained subtracting the model from the data and clearly shows the thin disk flaring.}
\label{fig:bmodel}
\end{figure*}
We cannot determine the moment when the (minor) merger(s) happened in FCC\,170, but we do not expect it to be recent because we do not see morphological disturbances in any of the maps (e.g. Fig.~\ref{fig:kin}).
In addition, direct mergers would be rare in the very dense environment where we observe FCC\,170 now \citep[e.g.][]{Dressler1984,Moore1996}.
However, they could have been rather probable in the proposed primordial (sub-)group, where FCC\,170 could have lived during the "pre-processing" phase before entering the cluster.

Although we have strong features characterizing each component, they also contain
a lower
mass fraction of
other populations (see Figs.~\ref{fig:agemet_bins}, \ref{fig:age_bins} and \ref{fig:met_bins}).
This mixing of the different populations, was also
probably caused by some merger activity subsequent to the accretion of stars, while we cannot rule out secular mixing processes. Possibly one or both mechanisms distributed the stars of different (younger) populations in the
different components, without erasing the global chemical differences.
Together with the fact that the accreted stars correspond only to a modest fraction of the stellar mass, their re-distribution 
could explain why we do not see any sign of accretion in the kinematics (e.g. we do not see any strong rotation lag).

This mixing culminates in the thin disk metallicity flaring, involving a sufficient amount of stars 
to make it clearly visible in the velocity dispersion and metallicity maps.
The thin-disk flaring was previously detected in the photometric analysis of FCC\,170 by \citet{Iodice2017b,Iodice2018}. 
A bi-dimensional model of the light distribution was obtained for each early-type galaxy in their sample.
The method, detailed in \citet{Iodice2016,Iodice2018}, consisted of extracting the azimuthally-averaged intensity profile, fitting isophotes in elliptical annuli with the same position angle and ellipticity.
They subtracted the bi-dimensional model to the observed light, obtaining a residual image showing all
deviations from pure ellipses in the shape of galaxies outskirts.
While for other galaxies in the sample they detected asymmetric elongations, a thin-disk flaring appeared in FCC\,170, as shown in Fig.~\ref{fig:bmodel} (and Fig.\,14 of \citealt{Iodice2018}).
Comparing the scales in Fig.~\ref{fig:bmodel} and Fig.~\ref{fig:pop}, we can confirm that the same flaring shape visible in light is reproduced by the chemical composition.
The flaring in the disk of FCC\,170 was also detected in the $g-i$ color map by
\citet{Iodice2018}.

Flarings have been related with mergers in several previous works
 \citep[e.g.][]{Walker1996,Kazantzidis2008,Villalobos2008,Read2009,Bournaud2009,Qu2011,Minchev2015}. 
\citet{Bournaud2009} claimed that the outer regions of disks were more sensitive to the 
kinematic heating caused by minor mergers, given their lower density. 
Also, in their simulations, the disruption of the satellite left 
most of the mass and energy to the outer disks.
In FCC\,170, the absence of any visible age radial gradient disfavors the inside-out formation scenario
proposed by \citet{Minchev2015}. We cannot rule out differences of few Gyr, though.
Flarings do not have to be necessarily related to mergers, they can also have their origin simply in the disk potential, as shown by \citet{Narayan2002}. However, the specific environment where FCC\,170 lives points to gravitational interactions as responsible for this flaring.
Because our galaxy mainly formed in a fast early process, mergers probably heated the outer thin disk when it was already formed, whereas the thick disk had formed before.
Our MUSE data are not deep enough to go sufficiently far from the midplane and see if the thick disk flares too.\looseness-2

\subsubsection{The boxy-bulge and the nuclear disk}
An X-shape in the metallicity map was previously observed in NGC\,4710 by \citet{Gonzalez2017} and in the Milky Way by \citet{Fragkoudi2018}.
\citet{Gonzalez2017} explained it with a kinematic fractionation model \citep{debattista2017}.
According to this scenario, stellar populations can be separated by an evolving bar according to 
their initial radial velocity dispersion (see also \citealt{Fragkoudi2017a}).
(Younger) more metal-rich stars would have been dynamically colder when the bar formed.
Therefore they would have been vertically redistributed in the central region of the bar in a way that produces the X-shaped structure.
More metal-poor stars, dynamically hotter, would have become vertically thicker in a more uniform (but weaker) box-shaped configuration.
An X shape in the age map and age vertical gradients were also predicted by \citet{debattista2017}.
They warned, though, that such age differences (of the order of 2\,Gyr) could be difficult to detect.
In fact, they were not observed in NGC\,4710 nor in FCC\,170. As also suggested by \citet{Gonzalez2017}, the age resolution might 
not have been sufficient to detect these differences in both galaxies.

To explain observations of a drop in velocity dispersion associated with nuclear disks, \citet{Emsellem2001} proposed a formation scenario from gas inflow along the bar.
Emission lines have been found in nuclear disks in other galaxies, supporting this picture \citep[e.g.][]{Chung2004}. 
On the one hand this scenario might not fit very well to FCC\,170, where we do not see any clear correspondence of age with kinematics (and metallicity) and no current star formation is detected.
Star formation finished relatively early in both nuclear disk and bar, if we do not
consider the minor younger population accreted later to the bar (Figs.~\ref{fig:agemet_hist_comp} and \ref{fig:agealpha_comp}). 
The nuclear disk must have formed almost
contemporarily to the bar. 
On the other hand, the nuclear disk looks colder than the bar, supporting the gas inflow scenario
and an \textit{in-situ} formation \citep{Morelli2010}.
It appears also more metal-rich than the bar, likewise the nuclear disks produced by \citet{Cole2014} with simulations of bar-driven gas inflows. 
Its lower values of [Mg/Fe] than the bar, suggest a slightly slower formation process.
It could have formed from fresh cold gas infalling, driven by the forming bar, with differences in timescales
that we are not able to detect in age.

\section{Summary and conclusions}
\label{conclusions}

We have observed a tiny nuclear disk, a prominent central X-shaped structure, a massive thin disk and a complex thick disk in the Fornax-cluster galaxy FCC\,170.
These structures are clearly distinguishable in the maps of velocity, velocity dispersion, skewness,
metallicity and [Mg/Fe].
This massive S0 galaxy looks overall old. Although some younger populations are present, 
no important differences are detected in the mean age of all the structural components.
Regarding the chemical composition, FCC\,170 is characterized by metal-rich thin disk, bar and nuclear disk, with solar or slightly supersolar [Mg/Fe] abundances.
On the other hand, it displays a more metal poor and $\alpha$-enhanced thick disk.
The star formation history indicates that the thick disk is the mix of different stellar populations.
A very old component contributes most of the mass. 
It has subsolar metallicities and larger [Mg/Fe] abundance than the rest of the galaxy.
A secondary younger component, even more metal-poor and $\alpha$-enhanced, formed around 10\,Gyr ago.
We have shown a certain level of mixing of the different populations throughout the galaxy.
A flaring of the dynamically colder thin disk clearly incorporates metal-rich stars to the geometrically defined thick disk region,
corresponding to a potential third component.

Supported by the results of our analysis, we propose a formation scenario for FCC\,170 including an
\textit{in-situ} formation followed by the accretion of \textit{ex-situ} stars.
We suggest that the star formation peaked early, between 13 and 14\,Gyr ago, 
probably accelerated by the "pre-processing"
environment (i.e. a preliminary group or sub-group) where the galaxy could have lived before its infall into the Fornax cluster.
Almost all the stellar mass was formed rather rapidly, as shown by its overall solar/supersolar [Mg/Fe] abundance.
On the one hand stars in the thick disk, the most $\alpha$-enhanced and metal-poor, would have been formed
in a shorter time and from gas not yet chemically enriched.
The star formation was probably a little slower
in the thin disk, bar and nuclear disk.
Here, the interstellar medium would have had time to evolve reaching supersolar metallicities.
The contribution by supernovae type-Ia would have become noticeable leading to abundance of [Mg/Fe] around solar.
Differences in the timescales would have been smaller enough not to be detected with our age resolution,
but
sufficient to change the dominant processes of
the gas enrichment.
Star formation would have been globally quenched after few Gyr, probably helped by gas stripping processes.
On the other hand, about 10\,Gyr ago, another population would have been formed on an even shorter timescale in an even more metal-poor satellite galaxy.
Afterwards, the satellite would have been accreted during a minor merger adding this younger population with a different chemistry to FCC\,170.
The same or other mergers could be also responsible for a relative mixing of the populations and for the thin-disk flaring.

This work is part of a more extended study on the origin of thick disks in external galaxies, setting them in a general context of galaxy formation and evolution. We aim to achieve a global view by gathering more complete samples:  including edge-on S0 galaxies in different regions of the Fornax cluster, but also adding field galaxies of different morphological types.

\begin{acknowledgements}
This work is based on observations collected at the European Organization for Astronomical Research in the Southern Hemisphere under ESO programme 296.B-5054(A).
The F3D team is grateful to the P.I.s of the Fornax Deep Survey (FDS)
with VST (R.F.\ Peletier and E.\ Iodice) who kindly provided the published
thumbnail of the FDS mosaic in the $r$ band centred on FCC~170 and of the 2D model. The authors are very grateful to Tom\'as Ruiz-Lara for providing his helpful STECKMAP tests.
FP acknowledges Fundaci\'on La Caixa for the financial support received in the form of a Ph.D. contract. FP, JFB, GvdV and RL 
acknowledge support from grant 
AYA2016-77237-C3-1-P from the Spanish Ministry of Economy and Competitiveness 
(MINECO).
EMC and LM, acknowledge financial support from Padua University through grants DOR1715817/17, DOR1885254/18 and BIRD164402/16.
RL acknowledges funding from the Natural Sciences and Engineering Research Council of Canada PDF award.
RMcD is the recipient of an Australian Research Council Future Fellowship (project number FT150100333).
IM acknowledges support by the Deutsche Forschungsgemeinschaft under the grant MI 2009/1-1.
GvdV acknowledges funding from the European Research Council (ERC) under the European Union's Horizon 2020 research and innovation programme under grant agreement No 724857 (Consolidator Grant ArcheoDyn).
\end{acknowledgements}





\bibliographystyle{aa}
\bibliography{biblio}
\appendix
\section{Uncertainties in stellar-kinematic and population parameters} \label{sec:unc}
\begin{figure*}
\centering
\resizebox{.51\textwidth}{!}
{
\includegraphics[scale=1,page=3,width=\textwidth]{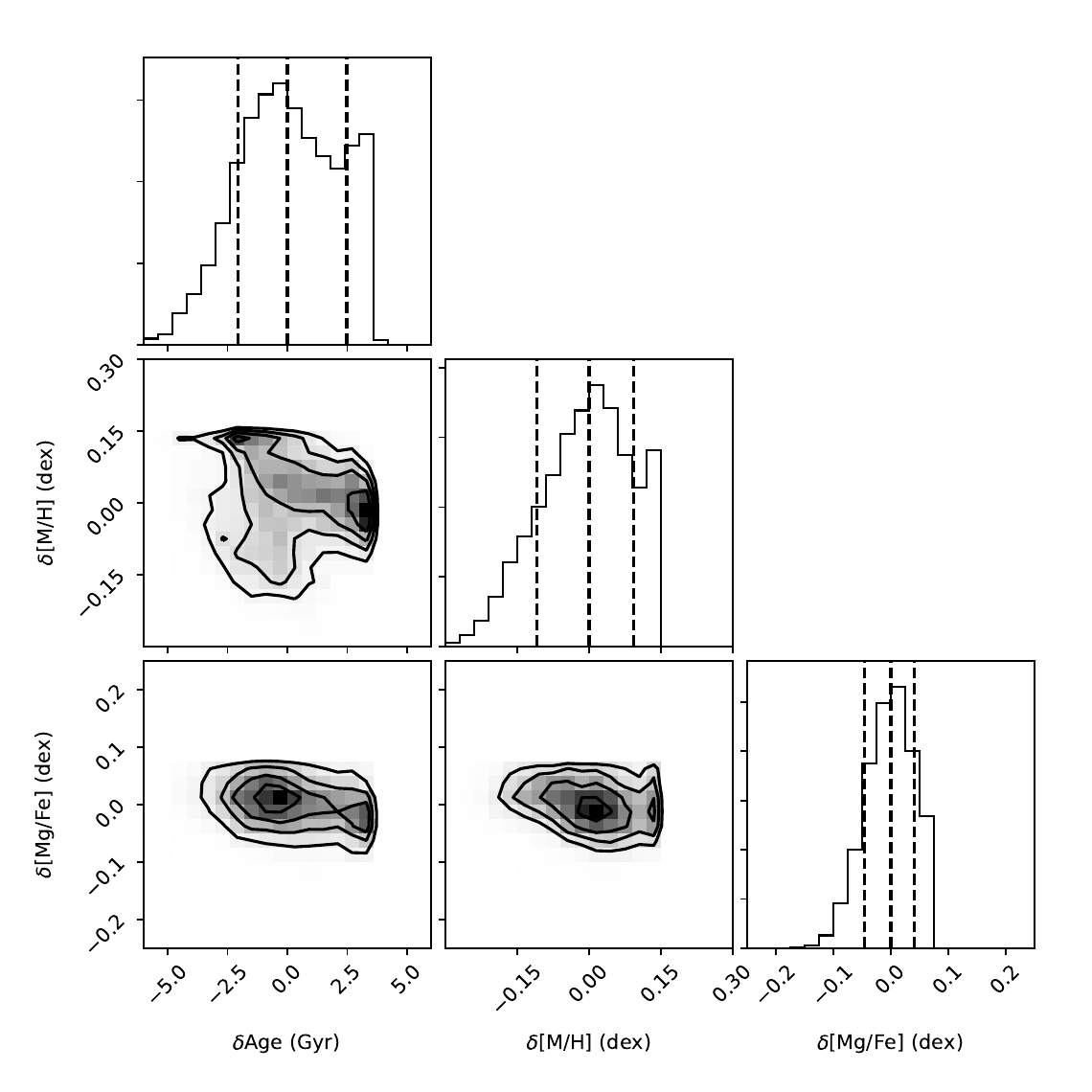}}
\caption{Corner plot of uncertainties of age, [M/H] and [Mg/Fe] for a spatial bin with SNR\,$\sim$\,40. The plotted $1\sigma$-uncertainty distributions result from fitting with pPXF 10\,000 Monte Carlo perturbations of the observed spectrum, with 2 different multiplicative polynomials and 3 different regularization parameters.
These uncertainties were calculated as the difference between the (weighted-average) individual values of age, [M/H] or [Mg/Fe] and the median value of all the 60\,000 fits. 
}
\label{fig:mc_unc}
\end{figure*}
We performed Monte Carlo simulations, as suggested by \citet{Cappellari2004}, to estimate the uncertainties on the stellar-kinematic and population parameters in Fig.~\ref{fig:kin} and \ref{fig:pop}.
We perturbed a series of spectra (of 10 spatial bins, 2 in each one of the 5 regions identified in Fig.~\ref{fig:morph}) with a Gaussian noise.
The noise level was estimated as the standard deviation of the residuals of a first pPXF fit (the residuals were calculated as the difference between the best fit and the original spectrum).
Each spectrum of the series was perturbed with 50 realizations.
We fitted each one of the perturbed spectra varying also the parameters of the fit. We used 5 different values of the degree of the multiplicative polynomial, from 6 to 10, and 14 different regularization parameters, from $10^{-2}$ to $10^{0.8}$.
Both the polynomials and the regularizations were selected around the values used for the analysis (for the regularization, up to the maximum allowed value according to \S\ref{sec:ppxf}).
This resulted in 3500 fits for each one of the 10 spatial bins.
The distribution of the 3500 results for the stellar-kinematic and population parameters allowed us to estimated their $1\sigma$ uncertainties.
The maximum uncertainties found among the 10 spatial bins were, for kinematics, about: 6\,km\,s$^{-1}$ for $V$, 9\,km\,s$^{-1}$ for $\sigma$, and 0.03 for $h_3$ and $h_4$.
These values were very similar to the estimates by \citet{Sarzi2018} for the galaxy FCC\,167 and the same target SNR\,=\,40.
The maximum uncertainties of the mean age, metallicity and [Mg/Fe] were respectively of 3\,Gyr, 0.1\,dex and 0.06\,dex.

Since these tests were too time consuming to increase the number of Monte Carlo realizations for the 10 selected spatial bins, we repeated the tests, now with 10\,000 perturbations, for only one of them (with SNR\,$\sim$\,40). In this case, we used only two values of the degree of the multiplicative polynomial: 8 (used in our analysis) and 1 (the minimum allowed when including regularization), and three levels of regularization: 0, 0.05 (the value used for the analysis, see \S\ref{sec:ppxf}) and 6.25 (maximum allowed value).
We confirmed that these new uncertainties had the same order of magnitude as our first estimates.
As an example, we show in the corner plot of Fig.~\ref{fig:mc_unc} the distributions of the $1\sigma$ uncertainties in age, [M/H] and [Mg/Fe]. 
These uncertainties were derived from the distribution of the 60\,000 results of fitting the 10\,000 perturbed spectra of this spatial bin. 
Each value in these uncertainty distributions was obtained as the difference between the result (i.e. the weighted average value of age, [M/H] or [Mg/Fe] from the specific fit) and the median value calculated among the 60\,000 results.

We consider that this method provides a good estimate but potentially still a lower limit because a white noise was added during the tests. The errors might be larger in real cases, in which the noise depends on wavelength. In addition to these statistical uncertainties, systematic uncertainties may be larger and very difficult to estimate. Factors which could be affecting them might be the imperfection of the SSP models and the fitting method, including IMF variations.
For these reasons, we adopt the maximum uncertainties presented in the last paragraph as the best estimates of errors in our method.

\section{Testing the age resolution in our method}\label{sec:tests}
\begin{figure*}
\centering
\resizebox{.51\textwidth}{!}
{
\includegraphics[scale=1,width=\textwidth]{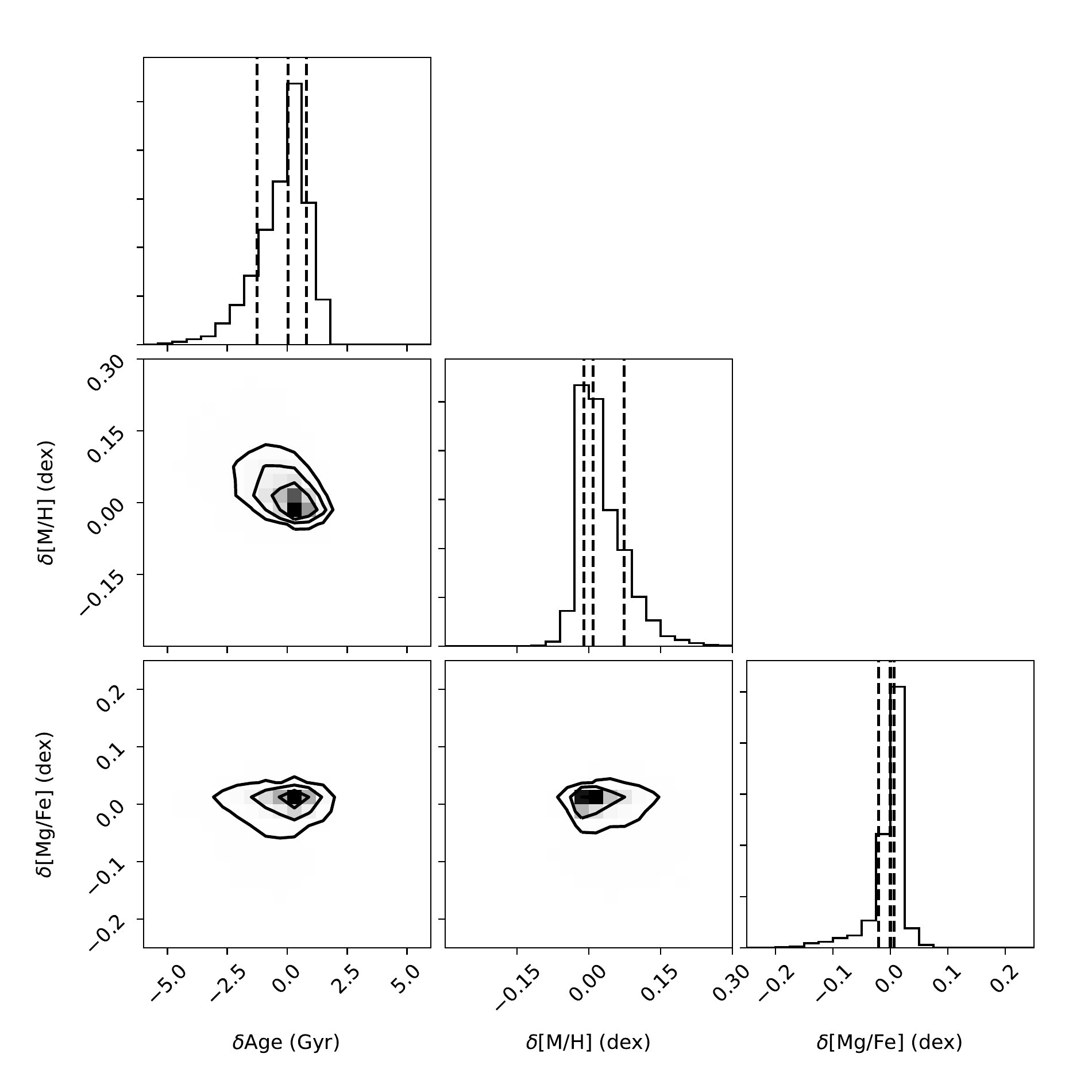}}
\caption{Corner plot of uncertainties of age, [M/H] and [Mg/Fe] from pPXF tests on mock spectra of SNR\,=\,40. 18\,000 different combinations of two SSPs of 12 and 14\,Gyr, with different metallicities and [Mg/Fe] abundance were included. A moderate value of the regularization parameter was used (0.05). The uncertainties were calculated as the difference between the mean values recovered from the fits and the true (mean) values. 
}
\label{fig:test_age_unc}
\end{figure*}
Aiming at testing the resolving power in age of our method, we performed a series of additional pPXF tests on recovering a simulated SFH consisting of two very old SSPs.
From the same set of models used for our analysis (see \S\ref{sec:ppxf}), we picked 12 and 14-Gyr-old SSPs of three different metallicities: [M/H]\,=\,-0.66, +0.06, and  +0.40\,dex, and two different values of [Mg/Fe]\,=\,0.00, and +0.40.
For each one of the 6 couples of metallicity and [Mg/Fe] values, we combined the 12 and 14-Gyr models in three different ways assigning them different weights (normalized to one): 0.25 for the youngest and 0.75 for the oldest, vice versa, and 0.5 for each SSPs. Therefore, we produced 18 simulated spectra in total.

We first fitted them with no added noise with pPXF. Three different levels of regularization (0, 0.05 and 6.25, see also Appendix~\ref{sec:unc}) and two degrees of the multiplicative polynomial (1 and 8) were used.
The simulated SFH, even the real fractions of the 12 and the 14-Gyr-old populations,  were perfectly recovered and were not affected by the different regularizations or polynomials.
Afterwards, each one of the 18 spectra was perturbed by 1000 Monte Carlo realizations, this time using noise levels corresponding to SNR\,=\,40 and 100.
We fitted these 36\,000 spectra with pPXF, while T. Ruiz-Lara (priv. comm.) used the code STECKMAP \citep{Ocvirk2006} for the same purpose. Taking into account the limited possibility of decreasing the smoothing in STECKMAP, the results were consistent with our pPXF tests described as follows.
With the two lowest regularizations, the resulting mean values of age, metallicity and [Mg/Fe] mimicked very well the real ones, for both levels of noise and all combinations of models.
We show, in Fig.~\ref{fig:test_age_unc}, a corner plot of the uncertainties in age, [M/H] and [Mg/Fe] estimated from the 18\,000 pPXF tests with SNR\,=\,40 and a moderate smoothing, i.e. the same level of regularization as in the analysis of MUSE data. These errors were calculated as the differences between the mean values from the fits and the true (mean) values. They are lower than the estimates in Appendix~\ref{sec:unc}.

For each combination of models, we calculated the average SFHs on the 1000 perturbations.
Only in few cases we recovered a clear distinction between the 12 and the 14-Gyr populations as different peaks (as expected, given the limited age sensitivity of spectral features in this very-old-age range).
In general, pPXF tended to give more weight to the oldest population than the real one.
Therefore, these tests confirmed the difficulty of distinguishing, with our method,  between very old populations with an age difference lower than our uncertainty estimate of Appendix~\ref{sec:unc}.
Several fits among the 1000 perturbations of each mock spectrum displayed in the SFHs spurious peaks at younger ages than the real ones, when using the two lowest regularization levels. 
However, these peaks were characterized by random age positions, so that they disappeared in the average SFHs.

\begin{figure}
\scalebox{0.9}{
\includegraphics[scale=0.46]{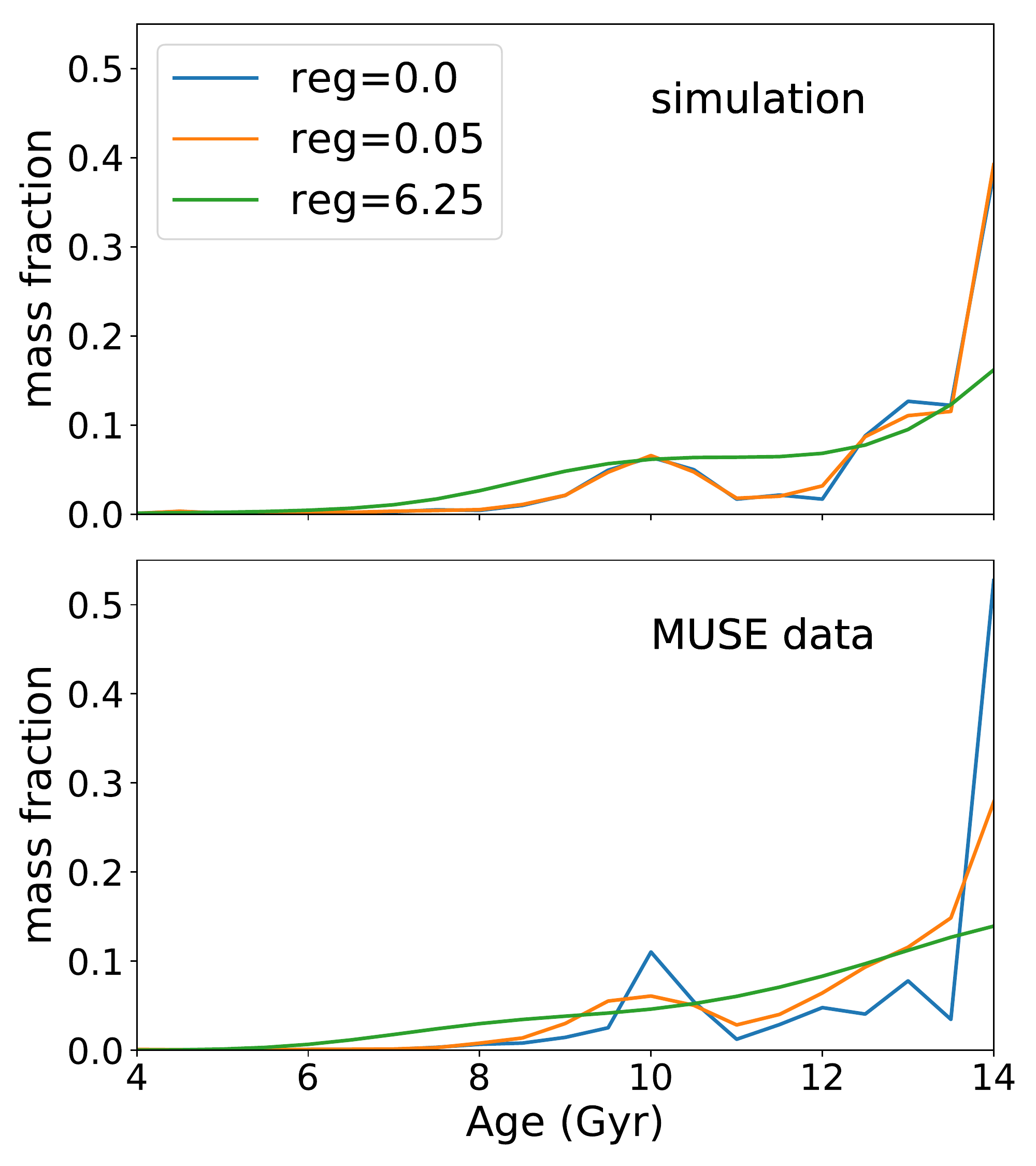}}
\caption{Comparison of the resulting SFH using three levels of regularization. Top panel: resulting average SFH from a mock spectrum, perturbed 1000 times with SNR\,=\,40, mimicking the SFH found in the thick disk (Fig.~\ref{fig:sfh_4c}). It is the composition of two 10 and 14-Gyr [Mg/Fe]-enhanced SSPs, of different metallicity. Bottom panel: average SFH from MUSE data of the thick-disk spatial bins. In both panels, three different regularizations were used.
}
\label{fig:test_comp}
\end{figure}

We can compare the 1000 perturbations to the different spatial bins in each structural component of FCC\,170, whose stellar populations are expected to be roughly the same in all bins.
Spatial bins in the inner thick disk, for instance, one of the regions with the lowest SNR, were characterized by very similar SFHs to each other, showing the stability of our results in comparison to the tests described here.
The SFH shown in Fig.~\ref{fig:sfh_4c} (second panel from bottom, but also bottom panel in Fig.~\ref{fig:test_comp}) was calculated averaging about 400 bins. Furthermore, it had the same shape if calculated from collapsing all the 400 spectra and reaching a much higher SNR (see also \S\ref{sec:SFH}).
The highest smoothing level in our tests led to a displacement of the mean age value towards younger ages, especially for SNR\,=\,40.
In our analysis, we kept regularization within a relatively low and safer regime.
Furthermore, the features in FCC\,170's SFH maintained their position with all tested regularizations (see Fig.~\ref{fig:test_comp}), in contrast to what happens when fitting the models, indicating the reliability of these results.

We repeated the tests, combining the 14-Gyr now with a 10-Gyr model, with the same combination of weights as with the 12-Gyr model, but now with fixed metallicities (respectively -0.25 and -0.66, roughly mimicking the two populations found in the thick disk of FCC\,170). Solar and enhanced [Mg/Fe] abundance were used for the 14-Gyr model. Perturbing 1000 times each one of the 6 mock spectra, we obtained 6000 fits. In all cases the two populations were recovered by pPXF as two different peaks, proportional to their weights.
In the top panel of Fig.~\ref{fig:test_comp} we show the average SFH (calculated on the 1000 fits) of one of the 6 combination of models: respective weights of 0.25 and 0.75 for the 10 and 14-Gyr populations, both [Mg/Fe]-enhanced.
We also compare this SFH with the one obtained from MUSE data for the thick disk of FCC\,170 (bottom panel). In each panel we plot three curves corresponding to the same three levels of regularization used in the previous set of tests. 
The figure shows that the peak at 10\,Gyr is smoothed with increasing regularization in both panels, staying however in the same age position.

\section{Bi-dimensional view of the SFH in FCC\,170}
\begin{figure*}
\centering
\resizebox{.74\textwidth}{!}
{
\includegraphics[scale=1,width=\textwidth]{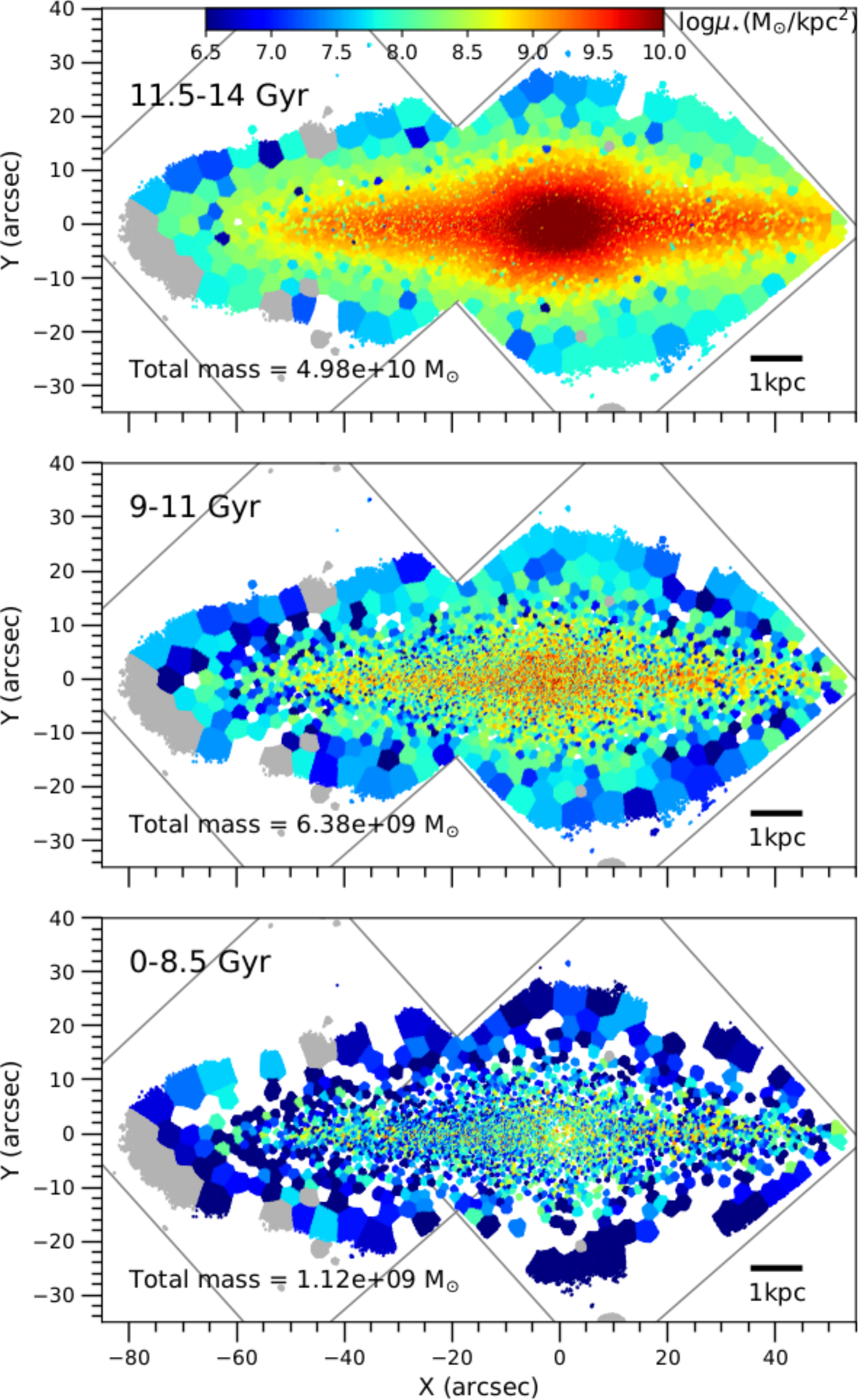}}
\caption{Two-dimensional view of the star formation history in FCC\,170. 
Stellar populations are mapped according to their age. Each map corresponds to a specific age bin: 
11.5\,--\,14\,Gyr in the top panel, 9\,--\,11\,Gyr in the middle panel and 0\,--\,8.5\,Gyr in the bottom panel. 
The color scale shows the mass density corresponding to the populations in the specific age 
bin.
The total mass in the age bin is indicated on bottom-left of each map.
The position of the two MUSE pointings is plotted in grey. A scale bar on bottom-right
of each map indicates the correspondence with physical units.
}
\label{fig:age_bins}
\end{figure*}

A two-dimensional view of the SFH is given by the maps in Fig.~\ref{fig:age_bins}.
Here, the stellar populations are divided into three age bins: 11.5\,--\,14\,Gyr, 9\,--\,11\,Gyr and 0\,--\,8.5\,Gyr. 
In each map, we have color coded the mass density contained in the corresponding age bin.
Populations more than 11.5\,Gyr-old (top panel of Fig.~\ref{fig:age_bins}) contribute about the 
87\,\%
of the mass. 
They are distributed everywhere, although the central region (box/peanut and nuclear disk) is clearly dominated 
by these very old populations.
The $\sim$\,10\,Gyr-old populations are plotted in the middle-panel map. They are present over all 
the galaxy, although mainly in the most massive structural components. 
Few stars are younger than 8.5\,Gyr since the mass contained in this age-bin populations 
is almost two orders of magnitude lower than the mass in very old stars. They are distributed mostly in the thin disk (bottom panel of Fig.~\ref{fig:age_bins}).

\section{Spatial distribution of metallicity in different ranges}
\begin{figure*}
\centering
\resizebox{.74\textwidth}{!}
{
\includegraphics[scale=1,width=\textwidth]{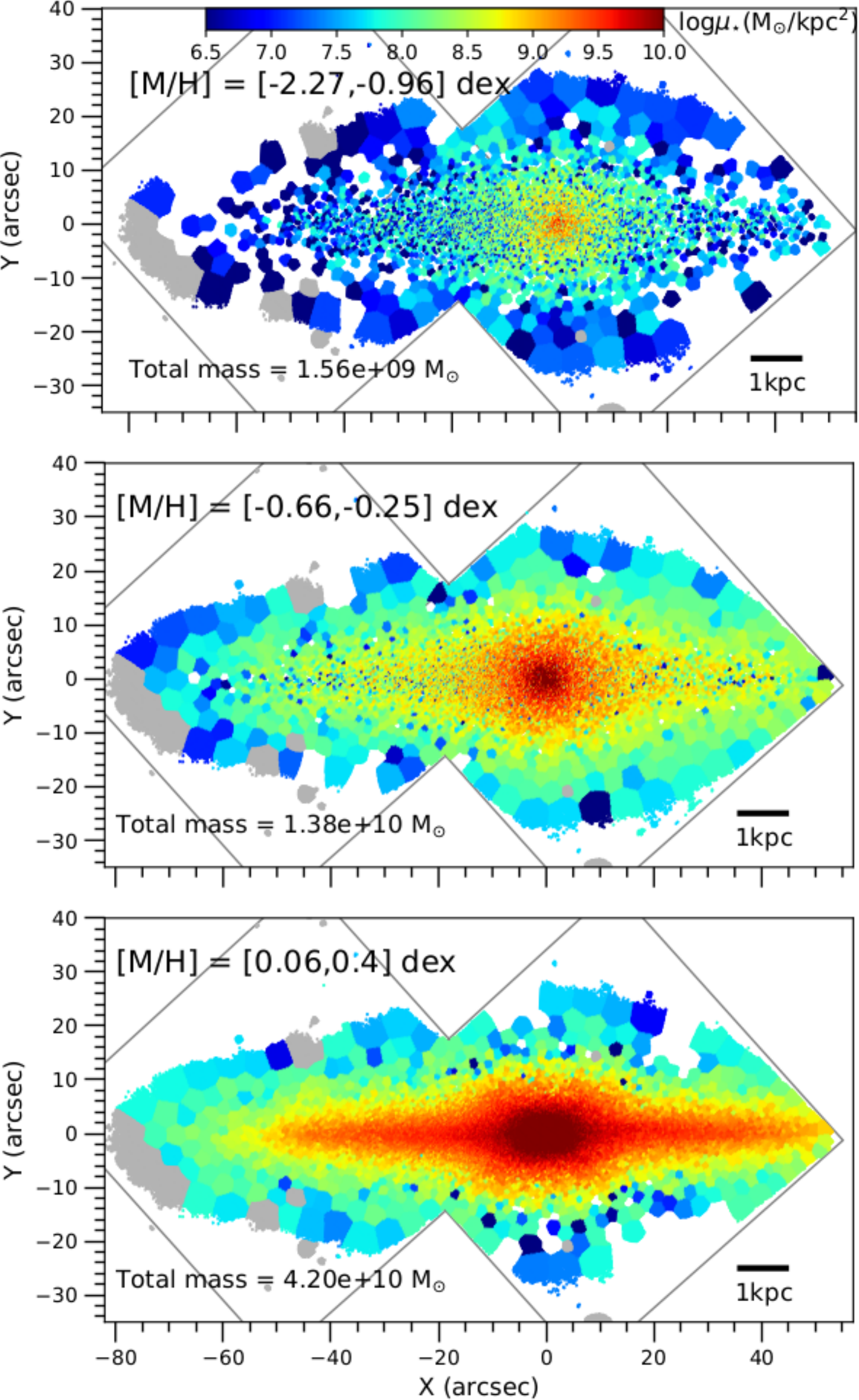}}
\caption{Maps of stellar populations in FCC\,170 with metallicities in three different bins: [M/H]\,=\,[-2.27,-0.96]\,dex in the top 
panel,
[-0.66,-0.25]\,dex in the middle panel and [0.06,0.4]\,dex in the bottom panel. 
The color scale shows the mass density corresponding to the populations in the specific metallicity 
bin.
The total mass in the metallicity bin is indicated on bottom-left of each map.
The position of the two MUSE pointings is plotted in grey. A scale bar on bottom-right
of each map indicates the correspondence with physical units.
}
\label{fig:met_bins}
\end{figure*}

Dividing our metallicity range into three bins, we can map the mass density of populations with different metallicities.
In the top panel of Fig.~\ref{fig:met_bins} we see the most metal poor stars.
These stars give the lowest contribution (about the 3\,--\,4\,\%) to the total mass. They are absent in numerous spatial bins of the outer regions.
Subsolar metallicities (top and middle panels) are present everywhere but are much denser in the central region, where a spherical symmetry is suggested. Populations in the supersolar metallicity range (bottom panel of Fig.~\ref{fig:met_bins}) contribute most of the mass. They follow the shape of the thin disk and the box/peanut, both dominated by these metal-rich stars.

\end{document}